\documentclass[aps,prd,twocolumn,10pt,superscriptaddress,nofootinbib,nobibnotes,longbibliography]{revtex4-1}

\usepackage{amssymb}
\usepackage{graphicx}
\usepackage{amsmath}
\usepackage{hyperref}
\usepackage{subfigure}
\usepackage{multirow}
\usepackage{setspace}
\usepackage{verbatim}
\usepackage{float}
\usepackage{color}
\usepackage{ulem}
\usepackage[utf8]{inputenc}

\begin{document}

\title{Gravitational waves from patterns of electroweak symmetry breaking: an effective perspective}

\author{Rong-Gen Cai}
\email{cairg@itp.ac.cn}
\affiliation{Institute of Fundamental Physics and Quantum Technology, Ningbo University, Ningbo, 315211, China}
\affiliation{CAS Key Laboratory of Theoretical Physics, Institute of Theoretical Physics, Chinese Academy of Sciences, Beijing 100190, China}
\affiliation{School of Physical Sciences, University of Chinese Academy of Sciences (UCAS), Beijing 100049, China}
\affiliation{School of Fundamental Physics and Mathematical Sciences, Hangzhou Institute for Advanced Study (HIAS), University of Chinese Academy of Sciences, Hangzhou 310024, China}
\affiliation{International Centre for Theoretical Physics Asia-Pacific, Beijing/Hangzhou, China}

\author{Katsuya Hashino}
\email{hashino@fukushima-nct.ac.jp}
\affiliation{Center for High Energy Physics, Peking University, Beijing 100871, China}
\affiliation{National Institute of Technology, Fukushima College, Nagao 30, Taira-Kamiarakawa, Iwaki, Fukushima 970–8034, Japan}

\author{Shao-Jiang Wang}
\email{schwang@itp.ac.cn}
\affiliation{CAS Key Laboratory of Theoretical Physics, Institute of Theoretical Physics, Chinese Academy of Sciences, Beijing 100190, China}
\affiliation{Asia Pacific Center for Theoretical Physics (APCTP), Pohang 37673, Korea}

\author{Jiang-Hao Yu}
\email{jhyu@itp.ac.cn}
\affiliation{CAS Key Laboratory of Theoretical Physics, Institute of Theoretical Physics, Chinese Academy of Sciences, Beijing 100190, China}
\affiliation{School of Physical Sciences, University of Chinese Academy of Sciences (UCAS), Beijing 100049, China}
\affiliation{School of Fundamental Physics and Mathematical Sciences, Hangzhou Institute for Advanced Study (HIAS), University of Chinese Academy of Sciences, Hangzhou 310024, China}
\affiliation{International Centre for Theoretical Physics Asia-Pacific, Beijing/Hangzhou, China}

\begin{abstract}
The future space-borne gravitational wave (GW) detectors would provide a promising probe for the new physics beyond the standard model that admits the first-order phase transitions. The predictions for the GW background vary sensitively among different concrete particle physics models but also share a large degeneracy in the model buildings, which motivates an effective model description on the phase transition based on different patterns of the electroweak symmetry breaking (EWSB). 
In this paper, using the scalar $N$-plet model as a demonstration, we propose an effective classification for three different patterns of EWSB: (1) radiative symmetry breaking with classical scale invariance, (2) Higgs mechanism in generic scalar extension, and (3) higher dimensional operators.
We conclude that a strong first-order phase transition could be realized for (1) and (2) with a small quartic coupling and a small isospin of an additional $N$-plet field for the light scalar field model with and without the classical scale invariance, and (3) with a large mixing coupling between scalar fields and a large isospin of the $N$-plet field for the heavy scalar field model.
\end{abstract}
\maketitle

\section{Introduction}\label{sec:introduction}

Despite of the success of the standard model (SM) of particle physics \cite{Zyla:2020zbs} as a low-energy effective field theory (EFT), it is incomplete in describing the puzzles of dark energy, dark matter (DM), cosmic inflation and baryon asymmetry of our Universe (BAU). The proposed solutions might call for a larger symmetry group for the ultraviolet (UV) completion, which should be broken into the SM symmetry group in our current epoch. Some of these symmetry breakings would trigger cosmic first-order phase transitions (FOPTs) (see \cite{Mazumdar:2018dfl} for a comprehensive review and \cite{Hindmarsh:2020hop} for a pedagogical lecture) proceeding by the bubble nucleations, bubble expansion and bubble collisions, which would generate a stochastic background of gravitational waves (GWs) (see \cite{Caprini:2015zlo,Caprini:2019egz} for recent reviews from LISA Collaboration \cite{Audley:2017drz} and \cite{Binetruy:2012ze,AmaroSeoane:2012je} for earlier reviews from eLISA/NGO mission \cite{AmaroSeoane:2012km}; see also \cite{Weir:2017wfa} for a brief review) transparent to our early Universe that is otherwise opaque to light for us to probe via electromagnetic waves if the 
FOPTs occurs before the recombination epoch. Therefore, the GWs detection serves as a promising and unique probe \cite{Figueroa:2018xtu,Axen:2018zvb} for the new physics \cite{Cai:2017cbj,Bian:2021ini} beyond SM (BSM) with FOPTs. 

Since the SM admits no FOPT but a cross-over transition due to a relatively heavy Higgs mass \cite{Kajantie:1996mn}, any model buildings with FOPTs should go beyond SM. However, a clean separation for BSM new physics with FOPTs from those without FOPTs turns out to be difficult, so does a clear classification for various FOPT models. Usually the FOPT models could be naively classified into models extended with higher-dimensional operators \cite{Zhang:1992fs,Bodeker:2004ws,Grojean:2004xa,Delaunay:2007wb,Huber:2007vva,Huber:2013kj,Konstandin:2014zta,Damgaard:2015con,Leitao:2015fmj,Harman:2015gif,Huang:2016odd,Balazs:2016yvi,deVries:2017ncy,Cai:2017tmh,Chala:2018ari,Dorsch:2018pat,deVries:2018tgs,Ellis:2019flb,Chala:2019rfk,Zhou:2019uzq,Phong:2020ybr,Kanemura:2020yyr,Hashino:2021qoq,Kanemura:2021fvp}, scalar singlet \cite{Profumo:2007wc,Espinosa:2011ax,Profumo:2014opa,Jinno:2015doa,Huang:2016cjm,Balazs:2016tbi,Curtin:2016urg,Hashino:2016xoj,Vaskonen:2016yiu,Kurup:2017dzf,Beniwal:2017eik,Kang:2017mkl,Chen:2017qcz,Chao:2017vrq,Beniwal:2018hyi,Shajiee:2018jdq,Alves:2018jsw,Grzadkowski:2018nbc,Hashino:2018wee,Ahriche:2018rao,Wan:2018udw,Chen:2019ebq,Alves:2019igs,Kannike:2019mzk,Chiang:2019oms,Kozaczuk:2019pet,Carena:2019une,Alves:2020bpi,DiBari:2020bvn,Pandey:2020hoq,Alanne:2020jwx,Paul:2020wbz,Xie:2020wzn,Kanemura:2022ozv}/doublet \cite{Huet:1995mm,Cline:1996mga,Fromme:2006cm,Cline:2011mm,Dorsch:2013wja,Dorsch:2014qja,Kakizaki:2015wua,Dorsch:2016nrg,Basler:2016obg,Bernon:2017jgv,Dorsch:2017nza,Huang:2017rzf,Basler:2017uxn,Barman:2019oda,Wang:2019pet,Zhou:2020xqi,Goncalves:2021egx,Graf:2021xku}/triplet \cite{Patel:2012pi,Inoue:2015pza,Blinov:2015sna,Chala:2018opy,Zhou:2018zli,Addazi:2019dqt,Benincasa:2019ejr,Brdar:2019fur,Paul:2019pgt,Bian:2019zpn,Niemi:2020hto,Wang:2020wrk,Borah:2020wut}/quadruplet \cite{Chala:2018ari}, composite Higgs \cite{Grojean:2004xa,Delaunay:2007wb,Grinstein:2008qi,Panico:2012uw,Grojean:2013qca,Csaki:2017eio,Espinosa:2011eu,Bian:2019kmg,DeCurtis:2019rxl,Xie:2020bkl,Chala:2016ykx,Fujikura:2018duw,Chala:2018qdf,Bruggisser:2018mrt,Bruggisser:2018mus}, supersymmetry (SUSY) \cite{Delepine:1996vn,Carena:1996wj,Apreda:2001tj,Laine:2012jy,Leitao:2012tx,Menon:2009mz,Curtin:2012aa,Carena:2012np,Cohen:2012zza,Liebler:2015ddv,Katz:2015uja,Pietroni:1992in,Davies:1996qn,Apreda:2001tj,Apreda:2001us,Menon:2004wv,Das:2009ue,Kozaczuk:2013fga,Kozaczuk:2014kva,Huber:2015znp,Demidov:2016wcv,Demidov:2017lzf,Bian:2017wfv,Akula:2017yfr,Georgi:1985nv,Cort:2013foa,Garcia-Pepin:2016hvs,Vega:2017gkk}, warp extra-dimensions \cite{Creminelli:2001th,Randall:2006py,Hassanain:2007js,Nardini:2007me,Konstandin:2010cd,Konstandin:2011dr,Servant:2014bla,Chen:2017cyc,Dillon:2017ctw,Bunk:2017fic,Marzola:2017jzl,Iso:2017uuu,vonHarling:2017yew,Megias:2018sxv,Fujikura:2019oyi,Agashe:2020lfz,Azatov:2020nbe,Bigazzi:2020phm,Megias:2020vek,Randall:2006py,Espinosa:2008kw,Iso:2009ss,Iso:2009nw,Konstandin:2010cd,Konstandin:2011dr,Okada:2014nea,Dorsch:2014qpa,Farzinnia:2014xia,Farzinnia:2014yqa,Jaeckel:2016jlh,Hashino:2016rvx,Jinno:2016knw,Kubo:2016kpb,Hashino:2016rvx,Marzola:2017jzl,Iso:2017uuu,vonHarling:2017yew,Chiang:2017zbz,Miura:2018dsy,Bruggisser:2018mrt,Bruggisser:2018mus,Brdar:2018num,Marzo:2018nov,Prokopec:2018tnq,Aoki:2019mlt,Bian:2019szo,Mohamadnejad:2019vzg,Kang:2020jeg,Chishtie:2020tze}, and dark/hidden sectors \cite{Espinosa:2008kw,Schwaller:2013hqa,Addazi:2016fbj,Hambye:2013sna,Jaeckel:2016jlh,Baker:2016xzo,Chala:2016ykx,Aoki:2017aws,Addazi:2017gpt,Tsumura:2017knk,Baldes:2017rcu,Baker:2017zwx,Bian:2018bxr,Breitbach:2018ddu,Baldes:2018emh,Croon:2018erz,Madge:2018gfl,Bian:2018mkl,Croon:2018new,Hall:2019rld,Fairbairn:2019xog,Katz:2016adq,Baldes:2017rcu,Long:2017rdo,Archer-Smith:2019gzq,Greljo:2019xan,Helmboldt:2019pan,Schwaller:2015tja,Coriano:2020kyb,Huang:2020bbe,Craig:2020jfv,Chao:2017ilw,Huang:2017laj,Addazi:2017nmg,Addazi:2017oge,Ayyar:2018ppa,Okada:2018xdh,Hashino:2018zsi,Hasegawa:2019amx,Hall:2019ank,Azatov:2019png,Haba:2019qol,Ghosh:2020ipy,Okada:2020vvb,Halverson:2020xpg,Dev:2016feu,DelleRose:2019pgi,vonHarling:2019gme,Croon:2019iuh,Dev:2019njv,Machado:2019xuc,Chiang:2020aui,Ghoshal:2020vud,Boeckel:2009ej,Schettler:2010dp,Boeckel:2011yj,Aoki:2017aws,Tsumura:2017knk,Capozziello:2018qjs,Khodadi:2018scn,Bai:2018vik,Bigazzi:2020avc,Huang:2020crf,Reichert:2021cvs}, which, however, are actually overlapping with each other when focusing on the sector that actually induces a FOPT. Nevertheless, most of the FOPT models could be regarded effectively as some kind of scalar extensions of the SM, while other FOPT models with fermion extensions \cite{Baldes:2018nel,Glioti:2018roy,Meade:2018saz,Matsedonskyi:2020mlz,Cao:2021yau}  are special on their own for triggering a PT, we therefore only focus on the scalar extensions of the SM.

On the other hand, some of the scalar extensions of the SM could be described and parametrized in the effective field theory (EFT) framework, in which the new particles are integrated out and only the SM degrees of freedom are kept.  
The EFT description was adopted before particularly for the higher-dimensional-operator extensions of the SM, which characterize the effect on the low-energy degrees of freedom when we integrate out the heavy degrees of freedom. However, the EFT description is only valid in the presence of a clear separation of scales, which is in conflict with the relatively low scale of the new degrees of freedom so as to introduce a large correction to the SM Higgs potential \cite{Postma:2020toi} in order to trigger a PT. Exceptions could be made for the Higgs-singlet extension with tree-level matching, though the EFT description is at most qualitative for dimension-six extension. Recently, a new perspective on this SM EFT description is made if the potential barrier separating the two minimums is generated radiatively instead of the tree-level barrier \cite{Camargo-Molina:2021zgz}. Nevertheless, we have found in this paper that the difficulty of an EFT description for the FOPT models could be circumvented by introducing a large number of scalar fields in the $N$-plet scalar field model. Therefore, for the electroweak phase transition process, the EFT description for some scalar extensions is not enough since in many cases the new light degree of freedom would contribute to the thermal plasma and thus one cannot integrate it out during phase transition. See also \cite{Gould:2019qek,Kainulainen:2019kyp,Niemi:2021qvp} for the dimensionally reduced effective field theory and its applications on reducing the uncertainties from the renormalisation scale dependence \cite{Croon:2020cgk,Gould:2021oba} and the thermal bubble nucleation calculation \cite{Gould:2021ccf,Lofgren:2021ogg,Hirvonen:2021zej}.

In this paper we instead take an intermediate strategy that lays between the specific new physics model and EFT treatment. We utilize a simplified model to illustrate feature of the electroweak phase transition (EWPT), which we call the effective model description on the phase transition. 
In this description, to capture different patterns of the EWPT and to compare the difference between new physics model and EFT description, we propose a specific effective model description: the general model extends the SM with an isospin $N$-plet scalar field, of which the light scalar case consists of a model with classical scale invariance (CSI) (model I) and a model without CSI (model II), while the heavy scalar case is simply a model with higher-dimensional operators, for example, a dimension-six operator (model III). 
The above cases could describe the patterns of the electroweak symmetry breaking (EWSB) via, for example, (1) radiative symmetry breaking, (2) Higgs mechanism,  and (3) EFT description of EWSB.
Our effective model description already covers those scalar models with $N$-plet on the market, such as (1) singlet models including a real scalar singlet extension of SM (xSM), composite Higgs model like SO(6)/SO(5) model, extra dimension model like radion model, dilaton model; (2) doublet models including SUSY model like minimal supersymmetry model (MSSM), two Higgs doublet model (2HDM), minimal dark matter model; (3) triplet models including left-right model, Type-II seesaw model. 
In other perspective, our effective model description consists of the simplified models (effective models I and II) for the realistic models (SUSY, composite Higgs, etc.) and an EFT model (model III). Therefore, our effective model provide an effective description for the EWSB that could admit a FOPT with associated GWs.

Although the effective scalar models describe different EWSB patterns, they share the same form of the Higgs potential. 
Thus, utilizing the polynomial potential form, we could analyse how the FOPT is realized in different cases. 
To show the source of a sizable barrier for realizing the first-order EWPT, we consider the following polynomial potential form in each of three models,
\begin{align}
V_p = C_2\phi^2+C_3\phi^3+C_4\phi^4+C_6\phi^6,
\end{align}
where $\phi$ is the order parameter in the effective potential, and $C_n$ are the effective couplings of $\phi^n$.
The $\phi^2$ and $\phi^4$ term can appear in it at tree-level, on the other hand, the $\phi^6$ term comes from high dimensional operator in the model (III).
The $\phi^3$ is the source of the first-order phase transition in the models (I) and (II), and it can be produced by the thermal loop effects.
Let us summarize the main features and also the main results of this work on realizing the FOPT in the forementioned three models:
\begin{itemize}
\item For model I, there are no massive parameters, and we consider the EWPT along a flat direction in the tree-level potential to avoid invalidating the perturbative analysis, and thus the potential form with finite temperature effects are roughly given by $V_p = C_2\phi^2+C_3\phi^3+C_4\phi^4$, where all terms are one-loop level effects coming from thermal loop effects and radiative corrections.
\item For model II, there are massive parameters in the Lagrangian unlike the model I. 
The potential of this model is still roughly given by $V_p = C_2\phi^2+C_3\phi^3+C_4\phi^4$ but the tree-level effects are now in $\phi^2$ and $\phi^4$ terms.
\item For model III, the high dimensional operator shows up in the potential $V_p = C_2\phi^2+C_3\phi^3+C_4\phi^4+C_6\phi^6$, where the tree-level effects are in $\phi^2$, $\phi^4$ and $\phi^6$ terms.
\end{itemize}
The main contributions to $C_n\phi^n$ in each of these models are summarized in Tab. \ref{TAB111}.
In the case of model I, not only $C_3$ term but also $C_2$ and $C_4$ terms are from loop level effects, and then the strongly first-order EWPT can be easily realized.
On the other hand, the model II receives tree-level effects from the $C_2$ and $C_4$ terms.
The source of barrier for the models I and II is the negative $C_3$ term from the thermal loop effects of bosons.
The model III has the high dimensional operator $C_6$, and thus we can have a negative $C_4$ term to generate a sizable barrier.
That is different point from models I and II.

 \begin{table}[tp]
  \label{TAB111}
  \centering
  \begin{tabular}{|c|c|c|c|c|}
    \hline
    Model & $C_2\phi^2$&$C_3\phi^3$&$C_4\phi^4$&$C_6\phi^6$\\
       \hline\hline
     I   & Loop & Loop & Loop & None\\
     \hline
     II & Tree & Loop & Tree & None \\
     \hline
      III & Tree & Loop & Tree & Tree \\
    \hline
    \end{tabular}
      \caption{The potential forms in the three types of the models where the EWSB occurs via (I) radiative symmetry breaking; (II) Higgs mechanism; (III) EFT description.
      Here, $\phi$ is order parameter, and $C_n$ is the effective coupling of $\phi^n$.
      The cubic term can be produced by thermal loop effects of bosons.
      To generate a sizable barrier, the cubic term will be negative in models I and II.
      In the model III, the quartic term can be negative to generate the barrier.}
\end{table}

The outline of this paper is as follows: in section \ref{sec:EFTmodel} we introduce our effective model description, whose effective potentials are detailed in section \ref{sec:Veff}. The resulted GWs from the FOPT models described above are extracted in a way depicted in section \ref{sec:GWPT}. The FOPT predictions are summarized in section \ref{sec:results}. The last section \ref{sec:condis} is devoted for conclusions and discussions. Appendix \ref{app:app1} is for some details on the model without CSI.

\section{Effective models}\label{sec:EFTmodel}

\begin{figure*}
\centering
\includegraphics[width=0.8\textwidth]{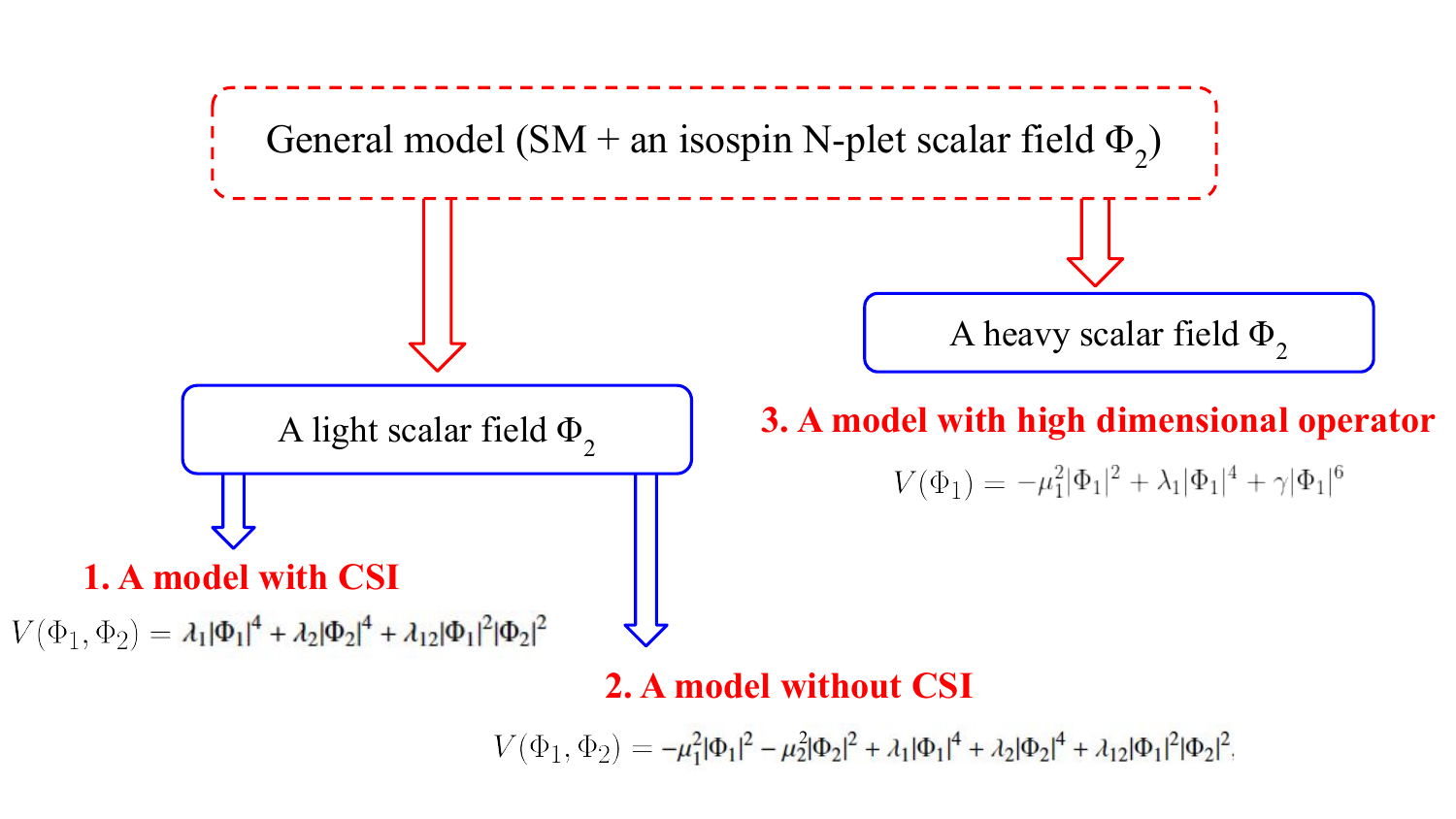}\\
\caption{Our effective model description for some BSM models with FOPTs from a general model with an additional scalar field, which is either light or heavy consisting of (I) the light scalar model with CSI, (II) the light scalar model without CSI, (III) the heavy scalar model with high dimensional operator.}
\label{fig:MODELS}
\end{figure*}

We focus on the model with an additional isospin $N$-plet scalar field $\Phi_2 \sim (I_{\Phi_2}, Y_{\Phi_2})$ charged under $SU(2)_I\times U(1)_Y$ gauge symmetry, where $I_{\Phi_2}$ is the isospin and $Y_{\Phi_2}$ is the hypercharge. The scalar boson fields in the model are given by 
\begin{align}
\Phi_1 = 
\left(\begin{array}{c} 
G^\pm,\\ \frac{h + iG^0}{\sqrt{2}}
\end{array} \right),
\quad
\Phi_2 = \frac{1}{\sqrt{2}} 
\left(\begin{array}{c} 
\phi_1 + i \phi_{1,i} \\
\phi_2 + i \phi_{2,i} \\
\vdots  \\
\phi_N + i \phi_{N, i} 
\end{array}\right),
\end{align}
where $\Phi_1$ is the SM-like double scalar field and $\phi_n$ ($\phi_{n,i}$) is the real (imaginary) part of the additional isospin $N$-plet scalar field with $N\equiv 2 I_{\Phi_2}+1$. These scalar bosons $\Phi_1$, $\Phi_2$ have classical fields: $\left\langle \Phi_1 \right\rangle/\sqrt{2}$ and $\left\langle \Phi_2 \right\rangle/\sqrt{2}$, which are related to the real part of the neutral scalar field. 
We will discuss the testability from the GW detection for three types of the extended models instead on different patterns of the EWSB.
These models are illustrated in Fig.~\ref{fig:MODELS}. 
Two types of them are the model with a light scalar field: (I) the model with classical scale invariance (CSI), (II) the model without CSI. The last type of the model is (III) the model with the $N$-plet scalar field with TeV scale. 
These models can realize the EWSB via (I) radiative symmetry breaking, (II) Higgs mechanism and (III) EFT description of EWSB, respectively.
For the simplicity of excluding the mixing terms, we assume $Z_2$ symmetry in such a way that the new scalar field is $Z_2$ odd while the others are $Z_2$ even.

\subsection{The model with classical scale invariance}

In the first type of the model, we impose CSI on the tree-level potential without any dimensional parameters, then the spontaneous EWSB is generated by radiative corrections~\cite{Coleman:1973jx} given later in \eqref{eq:Veff0CSI}. The Lagrangian of this model is
\begin{equation}
{\cal L} = {\cal L}_\mathrm{SM} + \left|D_\mu \Phi_2\right| ^2 - V_0(\Phi_1, \Phi_2),
\end{equation}
where $D_\mu = \partial_\mu  - i g T^a W_\mu^a -   i g' Y_{ \Phi_2} B_\mu/2$, $g'$ and $g$ are U(1)$_Y$ and SU(2)$_I$ gauge couplings, respectively, and $T^a$ is the matrix for the generator of SU(2)$_I$. The tree-level potential $V_0(\Phi_1, \Phi_2)$ is given by
\begin{equation}
\label{eq:CSItree}
V_0(\Phi_1, \Phi_2) = \lambda_1 |\Phi_1|^4+ \lambda_2 |\Phi_2|^4 +  \lambda_{12} |\Phi_1|^2 |\Phi_2|^2.
\end{equation}
If the isospin $I_{ \Phi_2}$ is 1/2, then there are some mixing terms in the potential, such as $|\Phi_1^\dagger \Phi_2|^2$. For simplicity, we neglect such terms in the potential. According to Refs.~\cite{Gildener:1976ih, Endo:2015ifa}, there may be a flat direction in the tree-level potential to assure the valid perturbative analysis. If not, the large logarithmic term may show up in the one-loop correction via the renormalization scale, for example, this scale in the $\phi^4$ theory is $Q\sim 3\lambda ve^{16\pi^2}$ \cite{Coleman:1973jx}. Therefore we assume a flat direction in the tree-level potential to avoid invalidating the perturbative analysis. The details of the effective potential will be discussed in Sec. \ref{sec:Veff}.

We mention here the constraints on this model. For simplicity, we assume that the additional scalar field $\Phi_2$ does not couple to the SM fermions and does not have the vacuum-expectation-value (VEV). On the other hand, the isospin $N$-plet scalar field $\Phi_2$ can interact with the SM gauge bosons. Then, the model is constrainted by the perturbative unitarity bound. According to Ref.~\cite{Hally:2012pu}, the isospin should be less than 7/2. We assume that the additional scalar field does not couple to the SM-like fermion in our strategy. Therefore, the typical collider constraints on our model is not much tight. For the neutral heavy Higgs, since it does not mix with the SM Higgs and does not couple to the SM fermions, the only experimental constraint comes from the vector boson scattering process, and thus it is very loose. For the charged Higgses, due to their degenerate masses to the neutral one, experimental constraints can be relaxed except for $h\gamma\gamma$ measurement. The additional charged scalar fields contribute to the Higgs coupling $h\gamma\gamma$~\cite{Earl:2013jsa}, which is given by $1.05\pm0.09$~\cite{ATLAS:2019nkf}. We may distinguish the models I and II by the results of measurements and observation of GW spectrum. In our work, we do not take into account the experimental constraints in the numerical analysis of the PT.

\subsection{The model without classical scale invariance}

In the model without CSI, there are mass parameters in the tree-level potential. The tree-level potential in this model is given as 
\begin{align}
\label{eq:generaltree}
V_0(\Phi_1, \Phi_2) = &- \mu_1^2|\Phi_1|^2 - \mu_2^2 |\Phi_2|^2  +  \lambda_{12} |\Phi_1|^2 |\Phi_2|^2\nonumber\\
&+ \lambda_1 |\Phi_1|^4+ \lambda_2 |\Phi_2|^4,
\end{align}
where $\mu_1^2>0$ and $\lambda_1, \lambda_2>0$ for the stability of the tree-level potential. The effective potential of this model will be given later in \eqref{eq:VeffwithoutCSI}.

\begin{figure}
\begin{center}
\includegraphics[width=0.5\textwidth]{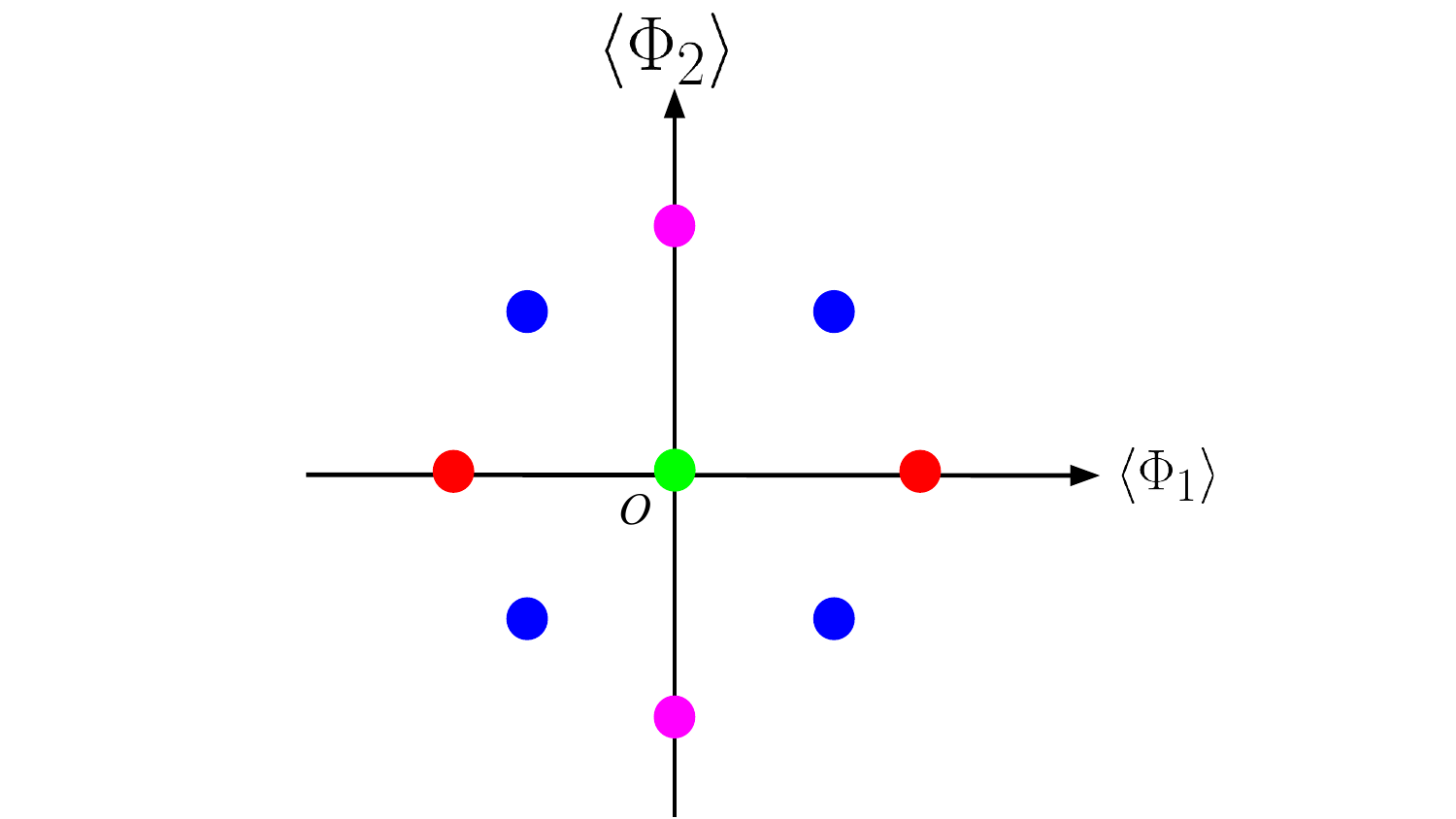}
\caption{ The extreme values of the tree-level potential of the model without CSI given by Eq.~\eqref{exte}.
In our analysis, the red (magenta) points will be global (local) minimum points and the blue points are saddle points.}
\label{fig:Vtree}
\end{center}
\end{figure}

Before discussing the effective potential with loop-corrections, we show the possible PT paths from the tree-level potential. At first, the extremal  values in the potential are obtained by 
\begin{equation}
\label{eq:eeee}
\frac{\partial V_0(\left\langle \Phi_1 \right\rangle,\left\langle \Phi_2 \right\rangle)}{\partial \left\langle \Phi_1 \right\rangle} = \frac{\partial V_0(\left\langle \Phi_1 \right\rangle,\left\langle \Phi_2 \right\rangle)}{\partial \left\langle \Phi_2 \right\rangle} = 0,
\end{equation}
which are solved by the following nine points in the field space as shown in Fig.~\ref{fig:Vtree},
\begin{align}
\label{exte}
&\left(\left\langle \Phi_1 \right\rangle,\left\langle \Phi_2 \right\rangle\right)  = (0, 0),\quad
\left(0, \pm \sqrt{\frac{\mu_2^2}{\lambda_2}}\right),\quad   
\left(\pm\sqrt{\frac{\mu_1^2}{\lambda_1}}, 0\right),\nonumber\\
&\left(\pm \sqrt{2\frac{\lambda_{12} \mu_2^2 - 2\lambda_2 \mu_1^2}{\lambda_{12}^2 - 4\lambda_1\lambda_2}} ,  \pm  \sqrt{2\frac{\lambda_{12} \mu_2^2 - 2\lambda_2 \mu_1^2}{\lambda_{12}^2 - 4\lambda_1\lambda_2}} \right), \nonumber\\
&\left( \pm  \sqrt{2\frac{\lambda_{12} \mu_2^2 - 2\lambda_2 \mu_1^2}{\lambda_{12}^2 - 4\lambda_1\lambda_2}} ,  \mp  \sqrt{2\frac{\lambda_{12} \mu_2^2 - 2\lambda_2 \mu_1^2}{\lambda_{12}^2 - 4\lambda_1\lambda_2}} \right).
\end{align}
The green point is the origin in the potential, the blue points are $\left\langle \Phi_2 \right\rangle\not=0$ at zero temperature and red and magenta points are along $\left\langle \Phi_1 \right\rangle$ and $\left\langle \Phi_2 \right\rangle$ axes, respectively. 
When the red or blue point is minimum, the scalar field $\Phi_1$ can have a VEV.
In this work, we especially focus on the one-step phase transition along $\left\langle \Phi_1 \right\rangle$ axis, because this path is the same as the CSI case. 
To realize the phase transition, we assume that the red point is global minimum and the blue point is not minimum.
We can assure such a situation by using the conditions for the determinants of the Hesse matrix and height of the potential.
The determinants of the Hesse matrix at the red, magenta and blue points are given by
\begin{align}
\label{hesseg}
\det[ {\cal H}_{red}] &=  \mu_1^2\left(  \frac{ \lambda_{12} \mu_1^2 }{\lambda_{1}} - 2 \mu_2^2\right),\\
\det[ {\cal H}_{mag}] &=  \mu_2^2\left(  \frac{ \lambda_{12} \mu_2^2 }{\lambda_{2}} - 2 \mu_1^2\right),\\
\det[ {\cal H}_{blue}] &=  -4  \frac{\left(\lambda_{12} \mu_2^2 - 2\lambda_2 \mu_1^2\right)\left(\lambda_{12} \mu_1^2 - 2\lambda_1 \mu_2^2\right)}{\lambda_{12}^2 - 4\lambda_1\lambda_2}.
\end{align}
Also, the height of the potential at these points are given by
\begin{align}
&V_0 \left(0,  \sqrt{\mu_2^2/\lambda_2}\right)= -\mu_2^4/4\lambda_2,\\
&V_0\left( \sqrt{\mu_1^2/\lambda_1}, 0\right)= -\mu_1^4/4\lambda_1\nonumber\\
&V_0 \left(   \sqrt{2\frac{\lambda_{12} \mu_2^2 - 2\lambda_2 \mu_1^2}{\lambda_{12}^2 - 4\lambda_1\lambda_2}} , \sqrt{2\frac{\lambda_{12} \mu_2^2 - 2\lambda_2 \mu_1^2}{\lambda_{12}^2 - 4\lambda_1\lambda_2}} \right)\nonumber\\
&= \frac{\lambda_1 \mu_2^4 + \lambda_2 \mu_1^4 - \lambda_{12} \mu_1^2\mu_2^2}{\lambda_{12}^2 - 4\lambda_1\lambda_2}.
\end{align}
The red point should be the lowest points among them.
From that, we can obtain the following conditions
\begin{align}
\label{con12}
 \frac{ \lambda_{12} \mu_1^2 }{\lambda_{1}} > 2 \mu_2^2,\quad   \frac{ \lambda_{12} \mu_2^2 }{\lambda_{2}} > 2 \mu_1^2,\quad \mu_2^4/\lambda_2< \mu_1^4/\lambda_1.
 \end{align}
Since the potential with these conditions have two minima at magenta and red points, two-step phase transition may be realized.
In our numerical analysis, we distinguish the phase transition pattern and focus on the one-step phase transition to compare differences of the results between the models with and without CSI. 

\subsection{The model with dimension-six operator from $\Phi_2$}

In the last type of model, $\Phi_2$ is assumed at the TeV scale and then we integrate out the additional scalar field. The tree-level potential in this model is given by
\begin{align}
\label{eq:generaltreeEFT}
V_0(\Phi_1, \Phi_2) &= - \mu_1^2|\Phi_1|^2 - \mu_2^2 |\Phi_2|^2  + \lambda_1 |\Phi_1|^4+ \lambda_2 |\Phi_2|^4 \nonumber\\
&+  \lambda_{12} |\Phi_1|^2 |\Phi_2|^2, 
\end{align}
where $\mu_2$ is at TeV scale. Otherwise it is the same as the model without CSI. Then, we integrate out the heavy $\Phi_2$ by loop-level matching~\cite{Henning:2014wua, Postma:2020toi} in our analysis, and the effective Lagrangian for the SM-like Higgs boson $h$ reads
\begin{equation}
\label{eq:EFTLag}
{\cal L}_\mathrm{EFT}^{(6)} \sim \frac{1}{2}|\partial_\mu h|^2 -\left(-\frac{1}{2}a_2 h^2 + \frac{1}{4}a_4h^4 + \frac{1}{6}a_6h^6  \right)
\end{equation}
with
\begin{align}
\label{eq:EFTcoup}
a_2 &= \mu_1^2,\\
a_4 &= \lambda_1 - (1+2I_{ \Phi_2})\frac{\lambda_{12}^2\mu_1^2}{9(4\pi)^2m_{\Phi_2}^2},\\
a_6 &=(1+2I_{ \Phi_2})\frac{1}{(4\pi)^2m_{\Phi_2}^2}\left(\frac{\lambda_{12}^3}{8}+\frac{\lambda_{12}^2\lambda_1}{6} \right),
\end{align}
where $m_{\Phi_2}^2= \mu_2^2 + \lambda_{12}v^2/2$ is the mass of the additional scalar field $\Phi_2$. The effective potential with radiative corrections of this model is given later in \eqref{eq:VeffD6withoutCSI}.

We note that Ref.~\cite{Postma:2020toi} suggests FOPT may be difficult in the model with a loop-level matching. According to their work, the FOPT requires the balancing between the dimension-four and dimension-six terms via $\frac{1}{4}\frac{U}{M^2}\frac{c_i\kappa^2}{16\pi^2}\sim \frac{m_h^2}{v^2}\sim0.12$ with $c_i$ = 1/2 (1) for a real (complex) scalar, where $U$ and $M$ correspond to the Higgs boson coupling to the heavy field and the mass parameter of heavy field, respectively. 
On the other hand, the parameter region for a valid EFT expansion requires $2c_{kin}v^2<1/2$ and $|a_8|v^2/a_6<1$, where $c_{kin}$, $a_6$ and $a_8$ are high dimensional operators for kinetic term, dimension 6 and 8 terms, respectively. Since the conditions also limit to $U/M^2$ and $c_i\kappa^2/16\pi^2$. Therefore, they concluded that the FOPT cannot be generated in the model.  At this time, we do not take into account other higher dimensional operators involving with the kinetic term.

\section{Effective potentials}\label{sec:Veff}

In this section, we discuss the forms of the effective potentials for our three effective model descriptions. To obtain the effective potential, we use the $\overline{\rm MS}$ scheme to absorb the divergence parts. Typically, the effective potential at one-loop level reads
\begin{align}
\label{eq:veff}
V_{\rm eff}&\left(\left\langle \Phi_1 \right\rangle,\left\langle \Phi_2 \right\rangle, T\right) = V_0 +\sum_i\frac{n_i}{64\pi^2} \, M^4_i\left(\left\langle \Phi_1 \right\rangle,\left\langle \Phi_2 \right\rangle\right)\nonumber\\
&\qquad\times\left(  \ln\left( \frac{M^2_i\left(\left\langle \Phi_1 \right\rangle,\left\langle \Phi_2 \right\rangle\right)}{Q^2} \right) - c_i \right) + \Delta V_T, 
\end{align}
where $M^2_i\left(\left\langle \Phi_1 \right\rangle,\left\langle \Phi_2 \right\rangle\right)$ is the field-dependent mass, $n_i$ is the number of the degree of freedom, $Q$ is the renormalization scale and $c_i$ is 3/2 ($i=$ boson, fermion) or 5/6 ($i=$ gauge boson). The one-loop thermal contribution to the potential~\cite{Dolan:1973qd} is 
\begin{align}
\label{eq:FINI}
\Delta V_T&= \frac{T^4}{2\pi^2}
\left\{ \sum_{i={\rm bosons}} n_i  \int_0^\infty d x\, x^2\right.\nonumber\\
&\times\left.\ln 
\left[ 1- \exp \left( -\sqrt{x^2+M_i^2(\langle\Phi_1\rangle, \langle\Phi_2\rangle, T)/T^2}\right) \right]\right.\nonumber\\
&\left. + \sum_{i = {\rm fermions}} n_i  \int_0^\infty d x\, x^2\right.\nonumber\\
&\times\left.\ln 
\left[ 1+ \exp \left( -\sqrt{x^2+M_i^2(\langle\Phi_1\rangle, \langle\Phi_2\rangle, T)/T^2}\right) \right] \right\}.
\end{align}
Here, we take into account the resummation effect $V_T^{\rm ring}$ obtained by~\cite{Carrington:1991hz}
\begin{align}
V_T^{\rm ring}=\frac{T}{12\pi}\sum_{i= \rm bosons}&n_i \left( (M_i^2(\langle\Phi_1\rangle, \langle\Phi_2\rangle, 0))^{3/2} \right.\nonumber\\
&\left.- (M_i^2(\langle\Phi_1\rangle, \langle\Phi_2\rangle, T))^{3/2}\right),
\end{align}
where the thermal mass $M_i^2(\langle\Phi_1\rangle, \langle\Phi_2\rangle, T)=M_i^2(\langle\Phi_1\rangle, \langle\Phi_2\rangle) + \Pi_i$ receives the thermal correction from the thermal self-energy $\Pi_i$. Finally, the effective potential with finite temperature effects is obtained by 
\begin{align}
\label{eq:ring}
V_{\rm eff}\left(\left\langle \Phi_1 \right\rangle,\left\langle \Phi_2 \right\rangle, T\right) = V_0 + V_{1-\rm loop} +  \Delta V_T + V_T^{\rm ring}.
\end{align}
The field-dependent masses in the effective potential of Eq.~\eqref{eq:veff} and thermal correction $\Pi_i$ in Eq.~\eqref{eq:ring} depend on the details of model. Therefore we will discuss the form of effective potential in each type of the model in the following.

\subsection{The model with classical scale invariance}

For the model \eqref{eq:CSItree} with CSI, the spontaneous EWSB occurs on the flat direction, which is assumed along $\langle\Phi_1\rangle$. Then, the effective potential is simply
\begin{align}
\label{eq:Veff0CSI}
V_{\rm eff}(\varphi, T=0)= A\varphi^4+B\varphi^4 \ln\frac{\varphi^2}{Q^2}, 
\end{align}
with $A$ and $B$ terms given by
\begin{align}
\label{eq:ABLOOP}
A= \sum_{i}  \frac{n_i}{64\pi^2 v^4} m_i^4 \left(\ln \frac{m_i^2}{v^2}-c_i\right), \quad
B= \sum_{i} \frac{n_i}{64\pi^2 v^4} m_i^4, 
\end{align}
where $m_i$ is the mass of the field species $i$ excluding the Higgs boson since this effect is at one-loop level in this model case. $n_i$ and $c_i$ in the potential are given by $(n_W,n_Z,n_t, n_{\Phi_2})=(6,3,-12, 2(2I_{\Phi_2}+1))$ and $(c_W,c_Z,c_t)=(5/6,5/6,3/2,3/2)$, respectively.
Using the stationary condition, we have 
\begin{align}
\label{eq:vevCSI}
\left. \frac{\partial V_{\rm eff}(\varphi, T=0)}{\partial \varphi}\right|_{\varphi=v}
= \ln\frac{v^2}{Q^2}+\frac{1}{2}+\frac{A}{B} = 0 ,
\end{align}
where the renormalization scale $Q$ is fixed. 
Because $A$ and $B$ are the loop effects, $v$ can be large in contrast to the case of $\phi^4$ theory. Also, the Higgs boson mass is obtained as
\begin{align}
\label{eq:mhCSI}
m_h^2 = \left. \frac{\partial^2V_{\rm eff}(\varphi, T=0)}{\partial \varphi^2}\right|_{\varphi=v}= 8 B v^2,
\end{align}
from which we can obtain the additional scalar boson mass $m_{\Phi_2}$ as
\begin{align}
\label{eq:MSCSI}
m_{\Phi_2} &=  \frac{C}{(2(2I_{\Phi_2}+1))^{1/4}},\\
C^4&\equiv 8\pi^2 v^2 m_h^2 -3m_Z^4 -6m_W^4 + 12m_t^4
\end{align}
with $C = 543\, \mathrm{GeV}$ determined by Eq.(10) of Ref.~\cite{Hashino:2015nxa}, and $\lambda_{12}$ coupling is obtained via $m_{\Phi_2}^2=\lambda_{12}v^2/2$ as 
\begin{align}
\label{eq:lam12CSI}
\lambda_{12} =  \frac{\sqrt{2}C^2}{v^2\sqrt{2I_{\Phi_2}+1}}.
\end{align}

With the help of Eqs.~(\ref{eq:vevCSI}) and (\ref{eq:mhCSI}), the one-loop effective potential at zero temperature could be obtained in terms of 
the renormalized mass of the Higgs boson as
\begin{align}
V_{\rm eff}(\varphi, T=0)
=\frac{m_h^2}{8v^2}\varphi^4\left(\ln\frac{\varphi^2}{v^2}-\frac{1}{2}\right).
\end{align}
Since the loop effects are renormalized into the Higgs boson mass, the value of $hhh$ coupling with one-loop effects does not depend on the model extension~\cite{Hashino:2015nxa}. The field-dependent masses and resummation effects in the effective potential with finite temperature effects are
\begin{align}
\label{withring1}
M^2_{\Phi_2}(\varphi,T)= &\frac{m^2_{\Phi_2}}{v^2}\varphi^2 + T^2\left((I_{\Phi_2}+1) \frac{\lambda_2}{6}+ \frac{m^2_{\Phi_2}}{3v^2} \right.\nonumber\\
&\left.+ I_{\Phi_2}(I_{\Phi_2}+1) \frac{g^2}{4} +Y_{\Phi_2}^2 \frac{g'^2}{16}  \right).
\end{align}
Similarly, the thermally corrected field-dependent masses of gauge bosons in the ($W^1, W^2, W^3, B$) basis are 
\begin{align} 
\label{withring2}
M_g^{2 (L, T)}(\varphi,T)&= \frac{\varphi^2}{4} 
\left(\begin{array}{cccccccc} 
g^2 &&& \\
&g^2 && \\
&&g^2&gg' \\
&&gg'&g'^2
\end{array}\right)\nonumber\\
&+ a_g^{(L,T)}T^2
\left(\begin{array}{cccccccc} 
 \pi_{W}  &&& \\
& \pi_{W}   && \\
&& \pi_{W} & \\
&&& \pi_{B} 
\end{array}\right),
\end{align}
where
\begin{align} 
\label{MWsq}
&M_W^2 =\frac{ g^2}{4}\varphi^2 ,\quad M_{WB}^2=\frac{ gg'}{4}\varphi^2 ,\quad M_{B}^2=\frac{ g'^2}{4}\varphi^2 ,\\
\label{piW}
& \pi_{W}  = \frac{g^2}{9}I_{ \Phi_2}\left( 1 + I_{ \Phi_2}\right) \left( 1 + 2I_{ \Phi_2}\right)  +\frac{11}{6}g^2,  \\
& \pi_{B}  =   \frac{g'^2 T^2}{12} \left( 1 + I_{ \Phi_2}\right)    Y_{ \Phi_2}^2   + \frac{11}{6}g'^2,\, a^L_g=1,\, a^T_T=0.
\end{align}
The field-dependent mass of fermion does not receive thermal correction in  $T^2$ so that
\begin{align} 
M_t^2 &=\frac{ m_t^2}{v^2}\varphi^2 .
\end{align}

In this model, there are three parameters in the tree-level potential,
\begin{align} 
\lambda_1, \lambda_2, \lambda_{12},
\end{align}
where $\lambda_1$ is zero in order to assume a flat direction along $\langle \Phi_1\rangle$ axis. According to Eq.~\eqref{eq:mhCSI}, the isospin number $I_{\Phi_2}$ is related to the mass $m_{\Phi_2}$.  Therefore, the free parameters in the effective potential with finite temperature effects are in fact
\begin{align} 
I_{\Phi_2}, Y_{ \Phi_2}, \lambda_2.
\end{align}

\subsection{The model without classical scale invariance}

For the model \eqref{eq:generaltree} without CSI, the effective potential is given by
\begin{align}
\label{eq:VeffwithoutCSI}
V_{\rm eff}\left(\varphi_1,\varphi_2, T\right) = V_0 +\sum_i\frac{n_i}{64\pi^2} \, M^4_i\,\left(  \ln \frac{M^2_i }{Q^2} - c_i \right) + \Delta V_T, 
\end{align}
where $\left\langle \Phi_1 \right\rangle=\varphi_1/\sqrt{2}$, $\left\langle \Phi_2 \right\rangle=\varphi_2/\sqrt{2}$ and
\begin{align}
\label{eq:V0withoutCSI}
V_0=- \frac{\mu_1^2}{2} \varphi_1^2 -\frac{\mu_2^2}{2} \varphi_2^2  + \frac{\lambda_1}{4} \varphi_1^4 +  \frac{\lambda_2}{4} \varphi_2^4  +   \frac{\lambda_{12}}{4} \varphi_1^2 \varphi^2_2.
\end{align}
There are the five model parameters in the tree-level potential and these parameters can be fixed in terms of following parameters,
\begin{align}
v, m_h, m_{\Phi_2}, \lambda_{12}, \lambda_2,
\end{align}
by the stationary conditions and the second derivatives of the effective potential. The details of them are given in Appendix \ref{app:app1}.

The field-dependent masses with resummation corrections from the finite temperature effects of the effective potential are
\begin{align}
&M^2_{h, \Phi_2}(\varphi_1,\varphi_2,T)=
  \frac{1}{2}\left(M_{11}^2+M_{22}^2 \right.\nonumber\\
  &\qquad\qquad\qquad\left.\mp 
    \sqrt{(M_{11}^2-M_{22}^2)^2 + 4 M_{12}^2 M_{21}^2} \right), \\
& M^2_{N_1}(\varphi_1,\varphi_2,T)=
  - \mu_1^2 +  \lambda_1\varphi_1^2 +   \frac{\lambda_{12}}{2} \varphi_2^2 \nonumber\\
  &+T^2\left( (2I_{ \Phi_2}+1) \frac{ \lambda_{12}}{12}  + \frac{\lambda_h}{4} + \frac{ 3g^2}{16}  + \frac{g'^2}{16} +\frac{y_t^2}{4} \right), \\
&  M^2_{N_2}(\varphi_1,\varphi_2,T)=
  - \mu_2^2 +  \lambda_2\varphi_2^2 +  \frac{\lambda_{12}}{2} \varphi_1^2 \\
  &+T^2\left( \frac{(I_{ \Phi_2}+1) \lambda_{2}}{6} +\frac{ \lambda_{12}}{6} + \frac{g^2}{4} \left( I_{ \Phi_2}^2 + I_{ \Phi_2}\right)    +  \frac{ g'^2 Y_{ \Phi_2}^2}{16}\right), \nonumber
\end{align}
where
\begin{widetext}
\begin{align}\label{eq:FmasshH}
& \begin{pmatrix}
    M_{11}^2 & M_{12}^2 \\
    M_{21}^2 & M_{22}^2
  \end{pmatrix}
               = 
  \begin{pmatrix}
    -\mu_1^2 + 3 \lambda_{1} \varphi_1^2  +  \frac{\lambda_{12}}{2}\varphi_2^2  &  \lambda_{12}  \varphi_1  \varphi_2 \\ 
 \lambda_{12}  \varphi_1  \varphi_2 & -\mu_2^2 + 3 \lambda_{2} \varphi_2^2  +  \frac{\lambda_{12}}{2}\varphi_1^2
  \end{pmatrix}\nonumber\\
&\quad+T^2
	\begin{pmatrix}
	 (2I_{ \Phi_2}+1) \frac{ \lambda_{12}}{12}  + \frac{\lambda_1}{4} + \frac{ 3g^2}{16}  + \frac{g'^2}{16} +\frac{y_t^2}{4} & 0 \\
	0 &  \frac{(I_{ \Phi_2}+1) \lambda_{2}}{6} +\frac{ \lambda_{12}}{6} + \frac{g^2}{4} \left( I_{ \Phi_2}^2 + I_{ \Phi_2}\right)    +  \frac{ g'^2Y_{ \Phi_2}^2}{16}
	\end{pmatrix}. 
\end{align}
\end{widetext}
Also the thermally corrected field-dependent masses of gauge bosons in the ($W^1, W^2, W^3, B$) basis are 
\begin{align} 
M_g^{2 (L, T)}&=  
\left(\begin{array}{cccccccc} 
M_W^2 &&& \\
&M_W^2 && \\
&&M_W^2&M_{WB}^2 \\
&&M_{WB}^2&M_B^2
\end{array}\right)\nonumber\\
&+ a_g^{(L,T)}T^2
\left(\begin{array}{cccccccc} 
\pi_{W}  &&& \\
& \pi_{W}   && \\
&& \pi_{W} & \\
&&& \pi_{B} 
\end{array}\right),
\end{align}
where
\begin{align}
\label{piWmassive}
M_W^2 &=\frac{ g^2}{4}\left(  \varphi_1^2  + I_W^2 \varphi_2^2 \right),\quad M_{WB}^2=\frac{ gg'}{4}\left(  \varphi_1^2  +Y_{ \Phi_2}^2 \varphi_2^2 \right),\\
 M_{B}^2&=\frac{ g'^2}{4}\left(  \varphi_1^2  + Y_{ \Phi_2}^2 \varphi_2^2 \right),\quad a^L_g=1,\quad a^T_T=0,\\
 \pi_{W}  &= \frac{g^2}{9}I_{ \Phi_2}\left( 1 + I_{ \Phi_2}\right) \left( 1 + 2I_{ \Phi_2}\right)  +\frac{11}{6}g^2,  \\
 \pi_{B}  &=   \frac{g'^2 T^2}{12} \left( 1 + I_{ \Phi_2}\right)    Y_{ \Phi_2}^2   + \frac{11}{6}g'^2.
\end{align}
In this model, the EWPT can be generated in multi-field space $(\varphi_1, \varphi_2)$. 
We will focus on the path of the phase transition along $ \varphi_1$ axis. Therefore, we will discuss the possibility of one-step and two-step PTs, and will especially focus on the path of the phase transition along $ \varphi_1$ axis to discuss the distinction among the three types of models with one-step PT.

\subsection{The model with dimension-six operator from $\Phi_2$}

For the effective model \eqref{eq:EFTLag} after integrating out the heavy scalar sector from \eqref{eq:generaltreeEFT}, the effective potential is 
\begin{align}
\label{eq:VeffD6withoutCSI}
V_{\rm eff}&\left(\varphi_1, T\right) = -\frac{1}{2}a_2 \varphi_1^2 + \frac{1}{4}a_4\varphi_1^4 + \frac{1}{6}a_6\varphi_1^6\nonumber\\
& +\sum_i\frac{n_i}{64\pi^2} \, M^4_i\,\left(  \ln\left( \frac{M^2_i }{Q^2} \right) - c_i \right) + \Delta V_T, 
\end{align}
where $a_2$, $a_4$ and $a_6$ are given in Eq.~(\ref{eq:EFTcoup}), and the field-dependent mass of Higgs boson along with its thermal correction are
\begin{align}
M_h^2(\varphi_1) &=-a_2   +3 a_4\varphi_1^2 + 5a_6\varphi_1^4,\\ \Pi_h &= T^2\left(  \frac{a_{4}}{4} + \frac{3g^2}{16}   +  \frac{ g'^2}{16} +   \frac{y_{t}^2}{4}  \right).
\end{align}
The thermally corrected field-dependent masses of gauge bosons in the ($W^1, W^2, W^3, B$) basis are 
\begin{align} 
M_g^{2 (L, T)}(\varphi,T)&= \frac{\varphi^2}{4} 
\left(\begin{array}{cccccccc} 
g^2 &&& \\
&g^2 && \\
&&g^2&gg' \\
&&gg'&g'^2
\end{array}\right)\nonumber\\
&+ a_g^{(L,T)}T^2
\left(\begin{array}{cccccccc} 
 \pi_{W}  &&& \\
& \pi_{W}   && \\
&& \pi_{W} & \\
&&& \pi_{B} 
\end{array}\right),
\end{align}
where
\begin{align} 
M_W^2 &=\frac{ g^2}{4}\varphi_1^2 ,\quad M_{WB}^2=\frac{ gg'}{4}\varphi_1^2 ,\quad M_{B}^2=\frac{ g'^2}{4}\varphi_1^2 ,\\
  \pi_{W} & = \frac{11}{6}g^2,\quad  \pi_{B} =    \frac{11}{6}g'^2,\quad a^L_g=1,\quad a^T_T=0
\end{align}
and the field-dependent mass of top quark is
\begin{align} 
M_t^2 &=\frac{ m_t^2}{v^2}\varphi_1^2 .
\end{align}
There are three new parameters in this potential, namely,
\begin{equation}
m_{\Phi_2}^2, \lambda_{12}, I_{ \Phi_2}.
\end{equation}

\section{Gravitational waves}\label{sec:GWPT}

The GWs from the FOPTs serve as the promising probe for the new physics of BSM, including our EFT models of EWSB. In this section, we briefly outline the computation procedures to carry out our model constraints.

\subsection{Phase-transition parameters}

For an effective potential $V_\mathrm{eff}$ exhibiting a false vacuum $\phi_+$ and a true vacuum $\phi_-$ separated by a potential barrier, a cosmological FOPT proceeds via stochastic nucleations of true vacuum bubbles followed by a rapid expansion until a successful percolation to fully complete the phase transition process. In this section, we describe the phase transition dynamics consisting of the bubble nucleation and bubble percolation, which could be determined by the thermodynamics of effective potential alone without reference to the microscopic physics that leads to the macroscopic hydrodynamics of bubble expansion.

\subsubsection*{Bounce action}

The bubble nucleations of true vacuum bubbles at finite temperature admit stochastic emergences of the field configuration $\phi(r)$ connecting a true vacuum region $\phi(r=0)\equiv\phi_0\lesssim\phi_-$ (assuming $\phi_+<\phi_-$) to the asymptotic false vacuum region $\phi(r\to\infty)\equiv\phi_+$ in an $O(4)$-symmetric manner $\mathrm{d}s^2=\mathrm{d}r^2+r^2\mathrm{d}\Omega_3^2$ or an $O(3)$-symmetric manner $\mathrm{d}s^2=\mathrm{d}\tau^2+\mathrm{d}r^2+r^2\mathrm{d}\Omega_2^2$ depending on their maximum value of the nucleation rates \cite{Coleman:1977py,Callan:1977pt,Linde:1980tt,Linde:1981zj}
\begin{align}
\Gamma(T)=\max\left[T^4\left(\frac{S_3}{2\pi T}\right)^\frac32e^{-\frac{S_3}{T}}, \frac{1}{R_0^4}\left(\frac{S_4}{2\pi}\right)e^{-S_4}\right],
\end{align}
where the $O(4)$ bounce action
\begin{align}
S_4=2\pi^2\int_0^\infty\mathrm{d}r\,r^3\left[\frac12\left(\frac{\mathrm{d}\phi_B}{\mathrm{d}r}\right)^2+V_\mathrm{eff}(\phi_B,T)-V_\mathrm{eff}(\phi_+,T)\right]
\end{align}
is evaluated at the solution $\phi_B$ of the equation-of-motion 
\begin{align}
\frac{\mathrm{d}^2\phi}{\mathrm{d}r^2}+\frac{3}{r}\frac{\mathrm{d}\phi}{\mathrm{d}r}=\frac{\partial V_\mathrm{eff}}{\partial\phi},\quad\phi'(0)=0,\quad \phi(\infty)=\phi_+,
\end{align}
while the $O(3)$ bounce action
\begin{align}
\frac{S_3}{T}=\frac{4\pi}{T}\int_0^\infty\mathrm{d}r\,r^2\left[\frac12\left(\frac{\mathrm{d}\phi_B}{\mathrm{d}r}\right)^2+V_\mathrm{eff}(\phi_B,T)-V_\mathrm{eff}(\phi_+,T)\right]
\end{align}
is evaluated at the solution $\phi_B$ of the equation-of-motion 
\begin{align}
\frac{\mathrm{d}^2\phi}{\mathrm{d}r^2}+\frac{2}{r}\frac{\mathrm{d}\phi}{\mathrm{d}r}=\frac{\partial V_\mathrm{eff}}{\partial \phi},\quad\phi'(0)=0,\quad\phi(\infty)=\phi_+.
\end{align}
In the realistic estimations, the vacuum decay rate from $S_4$ usually dominates over the thermal decay rate from $S_3/T$ at extremely low temperature when the potential barrier does not vanish even at $T=0$. $R_0$ is the bubble radius defined by $\phi_B(r=R_0)=(\phi_0-\phi_+)/2$.

\subsubsection*{Nucleation temperature}

During the whole process of bubble nucleation, the false vacuum becomes unstable once the temperature drops below the critical temperature defined by 
\begin{align}
V_\mathrm{eff}(\phi_+,T_c)=V_\mathrm{eff}(\phi_-(T_c),T_c).
\end{align}
However, the bubble nucleation is only possible when the temperature further drops down to $T=T_i<T_c$ defined
\begin{align}
\phi_B(r=0,T_i)=\phi_-(T_i)
\end{align}
due to the presence of the Hubble friction term in the bounce equation. Since then, one can count the number of nucleated bubbles in one Hubble volume during a given time elapse by
\begin{align}
N(T(t))=\int_{t_i}^{t}\mathrm{d}t\frac{\Gamma(t)}{H^3}=\int_{T}^{T_i}\frac{\mathrm{d}T}{T}\frac{\Gamma(T)}{H(T)^4},
\end{align}
until the nucleation temperature $T_n<T_i$ defined by the moment when there is exactly one bubble nucleated in one Hubble volume,
\begin{align}
N(T_n)=1.
\end{align}

\subsubsection*{Percolation temperature}

The progress of the phase transition could be described by the the expected volume fraction of the true-vacuum regions at time $t$ \cite{Guth:1982pn,Turner:1992tz,Ellis:2018mja},
\begin{align}
I(t)=\frac{4\pi}{3}\int_{t_i}^t\mathrm{d}t'\,\Gamma(t')a(t')^3r(t,t')^3,
\end{align}
where $r(t,t')$ is the comoving radius of a bubble at $t$ nucleated at an earlier time $t'$, 
\begin{align}
r(t,t')\equiv\frac{R_0}{a(t')}+\int_{t'}^t\frac{v_w(\tilde{t})\mathrm{d}\tilde{t}}{a(\tilde{t})}\approx v_w\int_{t'}^t\frac{\mathrm{d}\tilde{t}}{a(\tilde{t})}.
\end{align}
Here we omit the initial bubble radius $R_0$ and fix the time-dependent bubble wall velocity $v_w(t)$ at its terminal  $v_w$.  When the effects for the overlapping of true-vacuum bubbles and ``virtual bubbles" nucleated in the true-vacuum regions are taken into account, the fraction of regions that are still sitting at the false vacuum at time $t$ could be approximated by the exponentiation of $I(t)$,
\begin{align}
P(t)=e^{-I(t)}.
\end{align}
A percolation temperature $T_p<T_n$ is therefore defined by a conventional estimation \cite{Leitao:2012tx,Leitao:2015fmj,Ellis:2018mja,Ellis:2020awk,Wang:2020jrd}
\begin{align}
P(T_p)=1/e.
\end{align}
In what follows, the strength factor and characteristic length scale for determining the peak amplitude and frequency in the GW spectrum will be evaluated at the percolation temperature.

\subsubsection*{Strength factor}

To characterize the total released latent heat into the plasma, a parameter $\alpha$ is defined by \cite{Ellis:2018mja}
\begin{align}
\alpha(T)=\frac{\Delta\rho_\mathrm{vac}(T)}{\rho_\mathrm{rad}(T)},
\end{align}
with the total released vacuum energy defined by
\begin{align}
\Delta\rho_\mathrm{vac}(T)&=\rho(\phi_+,T)-\rho(\phi_-(T),T)\nonumber\\
&=\Delta V_\mathrm{eff}(T)-T\frac{\partial\Delta V_\mathrm{eff}(T)}{\partial T}
\end{align}
due to $\rho=\mathcal{F}+Ts=-p+T\frac{\partial p}{\partial T}=V_\mathrm{eff}-T\frac{\partial V_\mathrm{eff}}{\partial T}$ similar to the usual definition of latent heat $L(T_c)=\rho(\phi_+,T_c)-\rho(\phi_-(T_c),T_c)$ at critical temperature. Here $\Delta V_\mathrm{eff}(T)\equiv V_\mathrm{eff}(\phi_+(T),T)-V_\mathrm{eff}(\phi_-(T),T)$ for short.

\subsubsection*{Characteristic length scale}

The characteristic length scale $R_*$ for the peak position of the GW spectrum is estimated from the inverse duration $\beta$ of the phase transition via the average number density of bubbles as
\begin{align}
R_*=n_B^{-1/3}=(8\pi)^{1/3}v_w\beta^{-1},
\end{align}
where the inverse duration $\beta$ of the phase transition is computed by expanding the nucleation rate around the percolation time,
\begin{align}
\Gamma=e^{\beta(t-t_p)+\cdots}, &\quad P(t_p)=1/e, \\ \beta=\left.-\frac{\mathrm{d}}{\mathrm{d}t}\frac{S_3(t)}{T(t)}\right|_{t=t_p}&=\left.H(T)T\frac{\mathrm{d}}{\mathrm{d}T}\frac{S_3(T)}{T}\right|_{T=T_p}.
\end{align}

\subsection{Gravitational wave spectrum}

The GW spectrum from a FOPT consists of three contributions: the bubble wall collisions, the sound waves, and the MHD turbulences. The contribution from the MHD turbulences is usually sub-dominated and hence omitted in the current study. The contribution from the bubble wall collisions \cite{Kosowsky:1991ua,Kosowsky:1992rz,Kosowsky:1992vn,Kamionkowski:1993fg,Huber:2008hg} only dominates the total GW spectrum when the bubble walls collide with each other while they are still rapidly accelerating. In most of the cases, the runaway wall expansion is highly unlikely, and they usually collide with each other long after having approached to the terminal velocity. Therefore, we only consider the contribution from the sound waves, which is well fitted numerically by \cite{Hindmarsh:2013xza,Hindmarsh:2015qta,Hindmarsh:2017gnf,Cutting:2019zws}\cite{Ellis:2018mja,Guo:2020grp}
\begin{align}
h^2\Omega_\mathrm{sw}=8.5\times10^{-6}&\left(\frac{100}{g_\mathrm{dof}}\right)^\frac13\left(\frac{\kappa_v\alpha}{1+\alpha}\right)^2\left(\frac{H_*}{\beta}\right)v_w\nonumber\\&\frac{7^\frac72\left(f/f_\mathrm{sw}\right)^3}{\left(4+3\left(f/f_\mathrm{sw}\right)^2\right)^\frac72}\Upsilon(\tau_\mathrm{sw}),
\end{align}
with the peak frequency
\begin{align}
f_\mathrm{sw}=1.9\times10^{-5}\,\mathrm{Hz}\left(\frac{g_\mathrm{dof}}{100}\right)^\frac16\frac{T_\star}{100\mathrm{GeV}}\frac{1}{v_w}\frac{\beta}{H_*}.
\end{align}
Here the suppression factor $\Upsilon$ is only important for sufficiently long duration comparable to or even slightly larger than the Hubble time scale, which is also neglected in our current study.
Furthermore, both the wall velocity $v_w$ and efficiency factor of bulk fluid motions $\kappa_v$ are also set to be unity for simplicity.

To quickly locate the parameter space with a promising detectability of the GW signals, we will first use the ratio $\phi_C/T_C$ at the critical temperature as a roughly estimation for the size of the strength factor~\cite{Espinosa:2010hh,Caprini:2015zlo},
\begin{align}
\alpha\sim\alpha_\infty\simeq4.9\times10^{-3}\left(\frac{\phi_*}{T_*}\right)^2,
\end{align}
which is a typically reliable estimation for most of the non-supercooling electroweak PTs since there are no new relativistic degrees of freedom or new particles with couplings to the Higgs comparable to those of the $SU(2)_L$ gauge bosons or top quark.
To qualify the model detectability, the GWs signals $h^2\Omega_\mathrm{GW}(f)$ are compared to the sensitivity curves of some GW detectors $h^2\Omega_\mathrm{sen}(f)$ during the mission year $\mathcal{T}$ by the signal-to-noise ratio (SNR),
\begin{align}\label{eq:SNR}
\mathrm{SNR}=\sqrt{\mathcal{T}\int_{f_\mathrm{min}}^{f_\mathrm{max}}\mathrm{d}f\left[\frac{h^2\Omega_\mathrm{GW}(f)}{h^2\Omega_\mathrm{sen}(f)}\right]^2}.
\end{align}

\section{Summary of results}\label{sec:results}

\begin{figure*}
\centering
\includegraphics[width=0.8\textwidth]{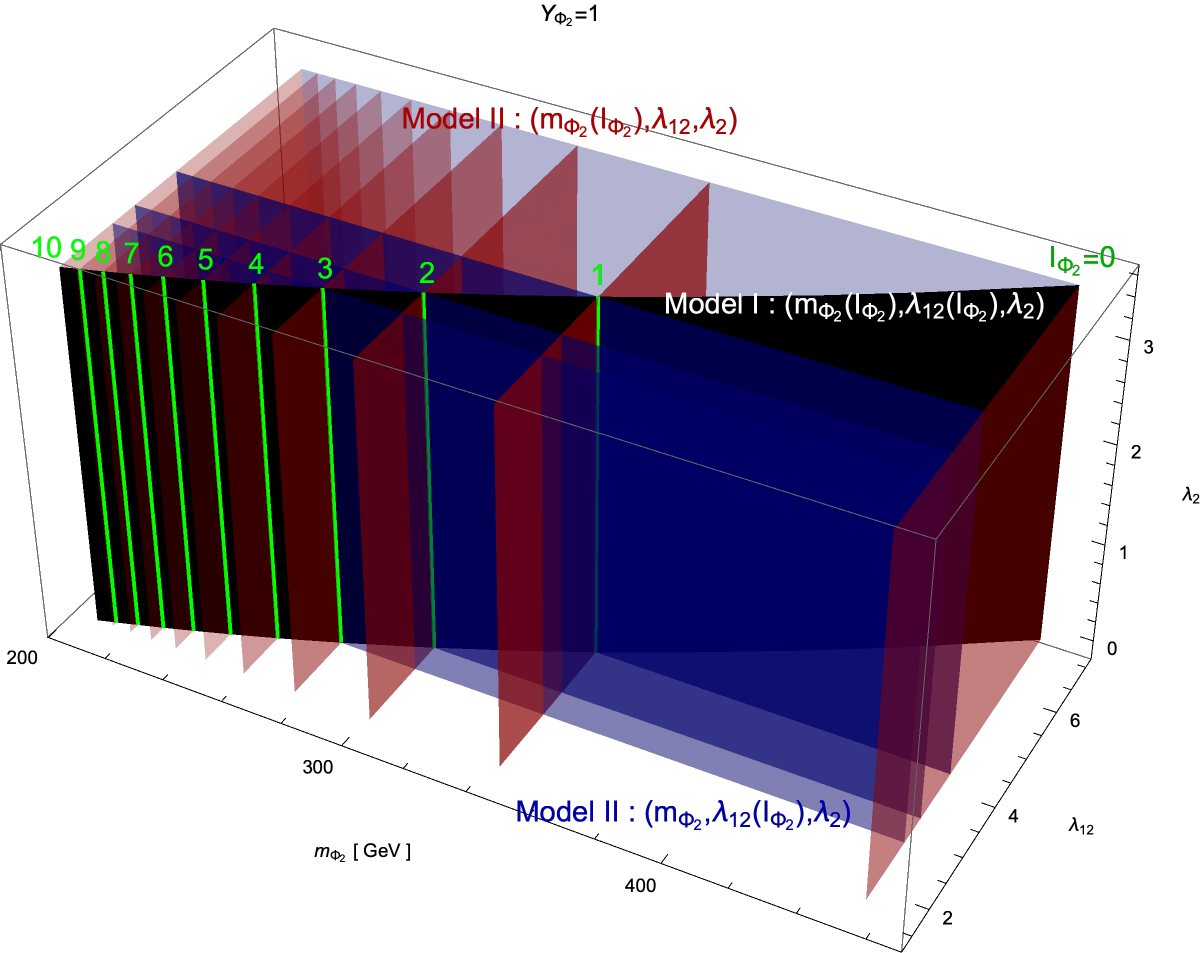}\\
\caption{The presentation for a part of the effective models of interest in the parameter space supported by $(m_{\Phi_2}, \lambda_{12}, \lambda_2)$ with a fixed $Y_{\Phi_2}=1$. The black strip is the model I with both $m_{\Phi_2}$ and $\lambda_{12}$ determined by $I_{\Phi_2}$ labeled in green. For model II, we take two illustrative examples with (1) the same $m_{\Phi_2}$ as the model I but a free $\lambda_{12}$ detached from $I_{\Phi_2}$ presented as red slices, and (2) the same $\lambda_{12}$ as the model I but a free $m_{\Phi_2}$ detached from $I_{\Phi_2}$ presented as blue slices. }
\label{fig:ModelSpace}
\end{figure*}

In this section, we show the results for the effective model descriptions with three types of the EWSB: (I) the light scalar model with CSI, (II) the light scalar model without CSI, and (III) the heavy scalar model with a higher dimensional operator after integrating out the additional heavy scalar. Besides the common SM parameters $(v, m_h, m_W, m_Z, m_t, g, g', \alpha_s, y_t)$, the new parameters in these three models are summarized below:
\begin{itemize}
\item For model I, there are five parameters ($I_{\Phi_2}, Y_{ \Phi_2}, \lambda_1, \lambda_2, \lambda_{12}$) from the effective potential \eqref{eq:Veff0CSI}.
By using the stationary condition, we can introduce the massive parameter EW vacuum by the dimensional transmutation~\cite{Coleman:1973jx}.
From the second derivative of the potential, the $\lambda_{12}$ is related to the SM effects, like Eq.~\eqref{eq:lam12CSI}. $\lambda_1$ is assumed to be zero for a flat direction along $\langle\Phi_1\rangle$ axis. 
Eventually, there are three free dimensionless parameters: $(I_{\Phi_2}, Y_{ \Phi_2}, \lambda_2)$.

\item For model II, there are seven parameters $(I_{\Phi_2}, Y_{ \Phi_2}, \mu_1, \mu_2, \lambda_1, \lambda_2, \lambda_{12})$ in the effective potential.
With the stationary condition, the massive parameter $\mu_1$ can be replaced by the EW VEV $v$.
From the second derivatives of the potential, $\lambda_1$ and $\mu_2$ parameters are given by the Higgs boson mass $m_h$ and the additional scalar boson mass $m_{\Phi_2}$.
Eventually, there are five free parameters: $( I_{\Phi_2}, Y_{ \Phi_2}, \lambda_2, m_{\Phi_2}, \lambda_{12})$. 
\item For model III, there are five parameters $(I_{\Phi_2}, \mu_1, m_{\Phi_2}, \lambda_1, \lambda_{12} )$ from the tree-level potential \eqref{eq:EFTLag}, two of which, for example, $\mu_1$ and $ \lambda_1$, could be fixed by the EW vacuum normalization conditions in terms of the EW VEV $v$ and the Higgs mass $m_h$, leaving behind three free parameters $( I_{\Phi_2}, m_{\Phi_2}, \lambda_{12})$.
\end{itemize}

Note that in the models I and II, although the hypercharge $Y_{\Phi_2}$ is a free parameter, it does not significantly change the PT results since its effect is proportional to $(m_W-m_Z)$ via Daisy diagram contributions. Thus, we take $Y_{\Phi_2}$ = 1 as an illustrative example in the following PT analysis. For consistent comparison, we consider the models with common values for the shared free parameters. For example, the model I and model II can be compared in the parameter space of $ I_{\Phi_2}$ and $\lambda_2$ for given choices of $m_{\Phi_2}$ and $\lambda_{12}$ as illustrated in Fig. \ref{fig:ModelSpace}.
The black plane in Fig.~\ref{fig:ModelSpace} corresponds to the parameter region for model I.
The $m_{\Phi_2}$ is proportional to the $\lambda_{12}$ in this model.
Such a relation is given by a line in the plane of ($m_{\Phi_2}$, $\lambda_{12}$.
Since the $\lambda_2$, which corresponds to the z axis in this figure, is independent free parameter, the black plane in Fig.~\ref{fig:ModelSpace} corresponds to the parameter region for model I.
The green lines are the $I_{\Phi_2}$ in the model I.
Since the model II has five free parameters, the all free parameters in this model cannot be determined, even if the parameters have the same values as the CSI model.
Such parameter regions corresponds to the red and blue planes.

\subsection{The model with classical scale invariance}

\begin{figure*}
\begin{center}
\includegraphics[width=0.4\textwidth]{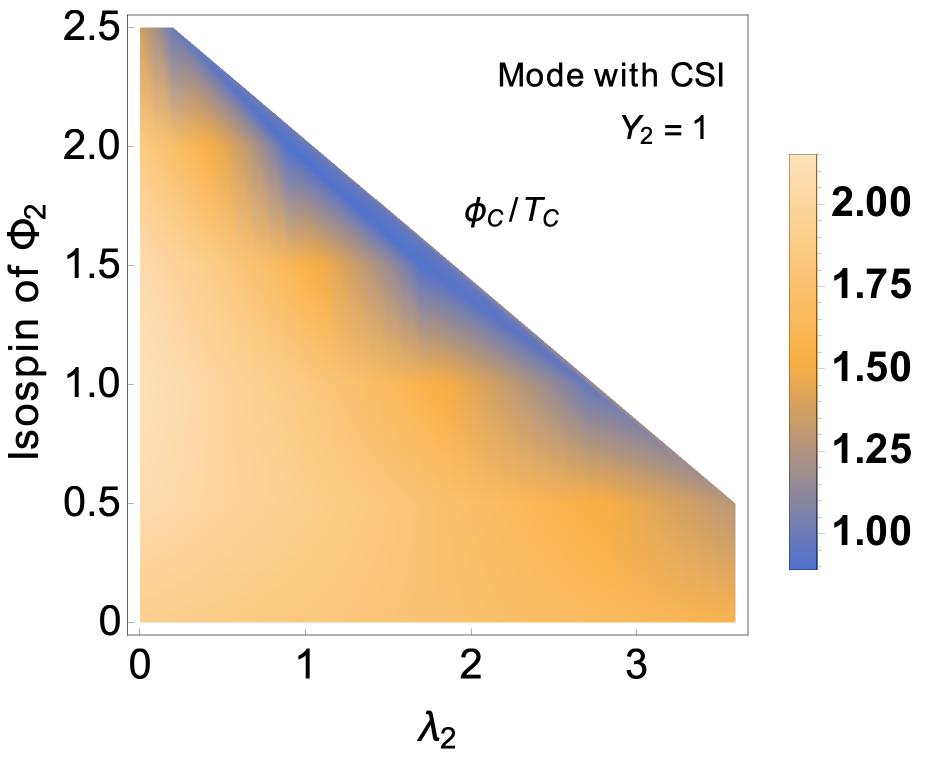}
\includegraphics[width=0.4\textwidth]{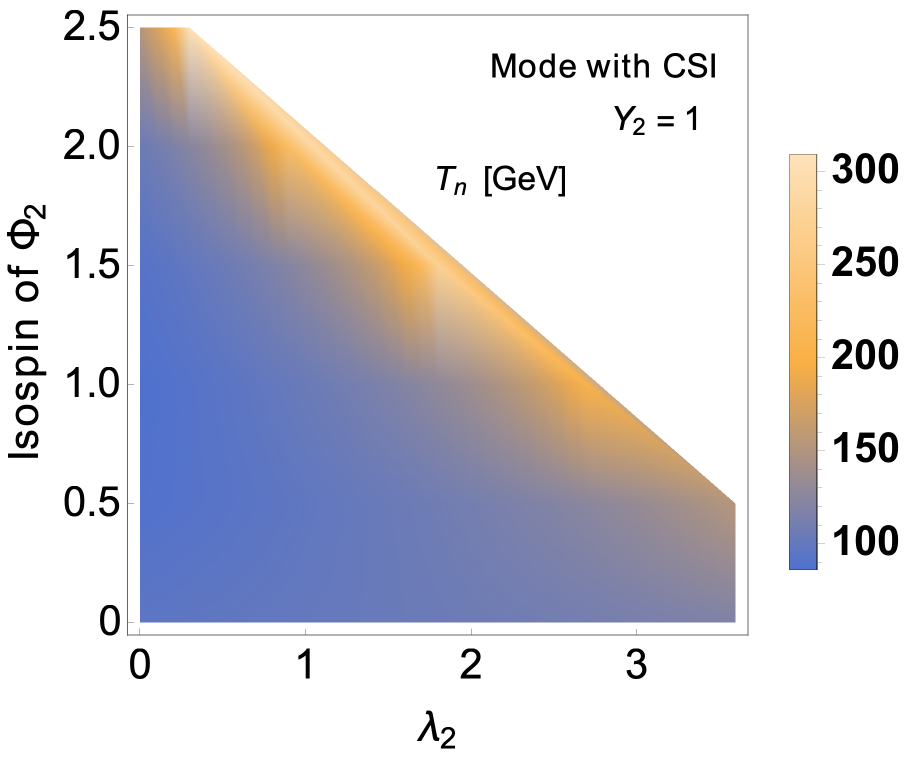}\\
\includegraphics[width=0.4\textwidth]{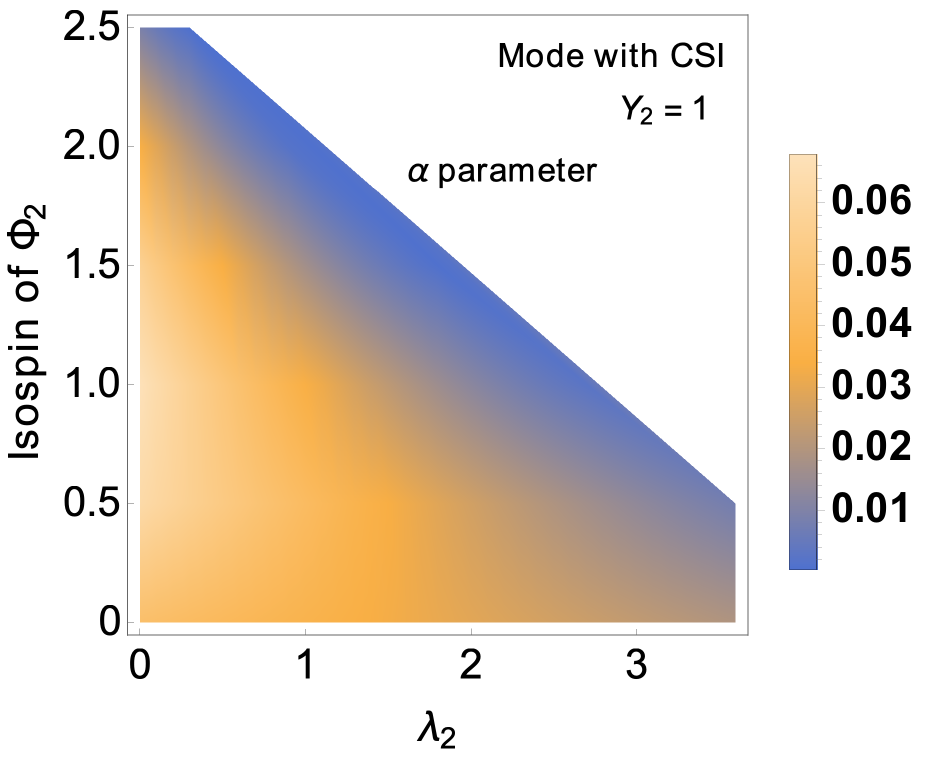}
\includegraphics[width=0.4\textwidth]{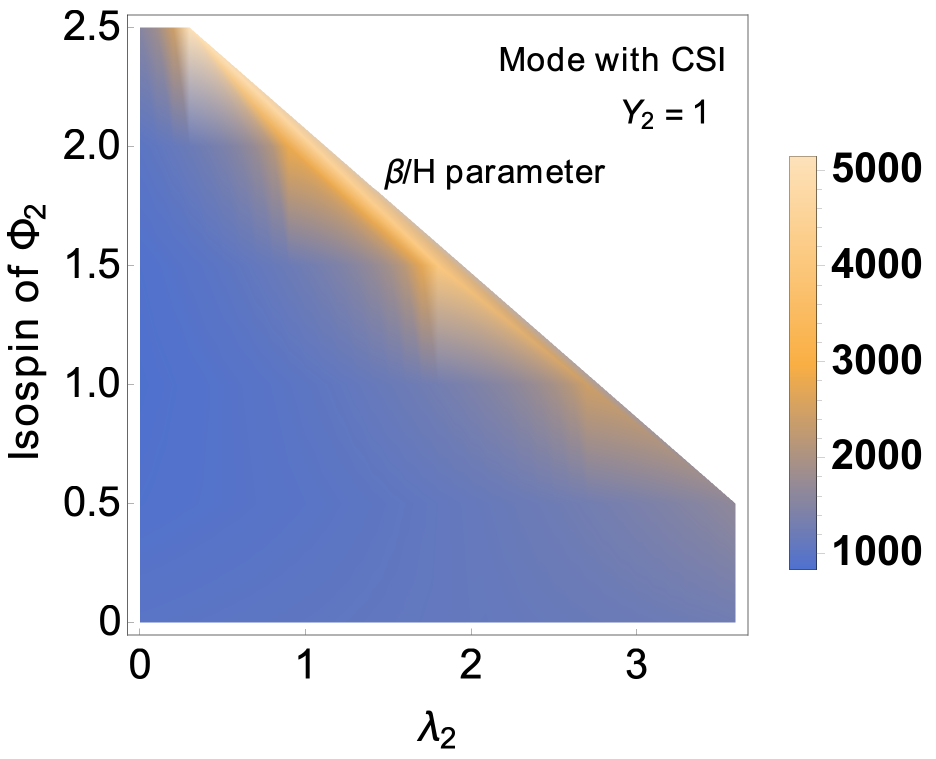}\\
\caption{The values of $\phi_C/T_C$, $T_n$, $\alpha$ and $\beta/H$ for the model I with respect to $\lambda_2$ coupling and isospin $I_{ \Phi_2}$ of the additional scalar boson fields.} 
\label{fig:PTGWCSI}
\end{center}
\end{figure*}

\begin{figure*}
\begin{center}
\includegraphics[width=0.3\textwidth]{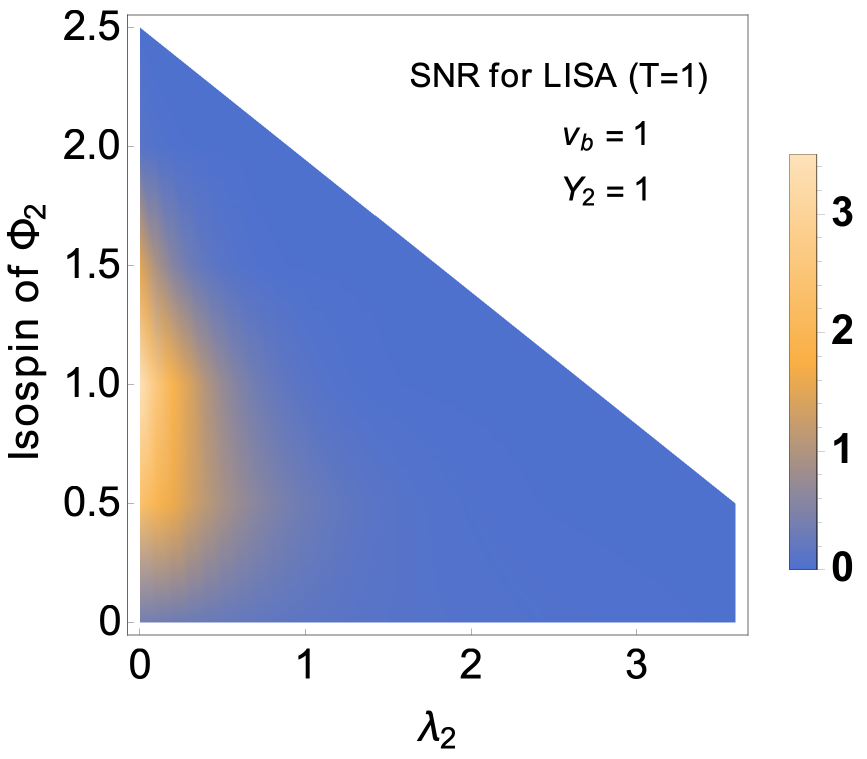}
\includegraphics[width=0.3\textwidth]{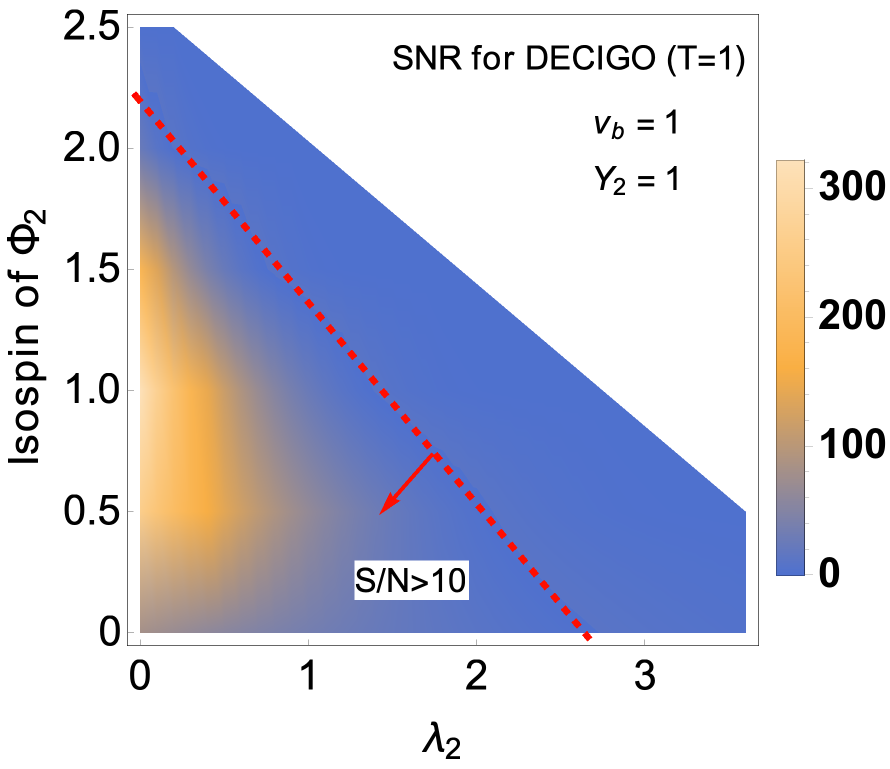}
\includegraphics[width=0.3\textwidth]{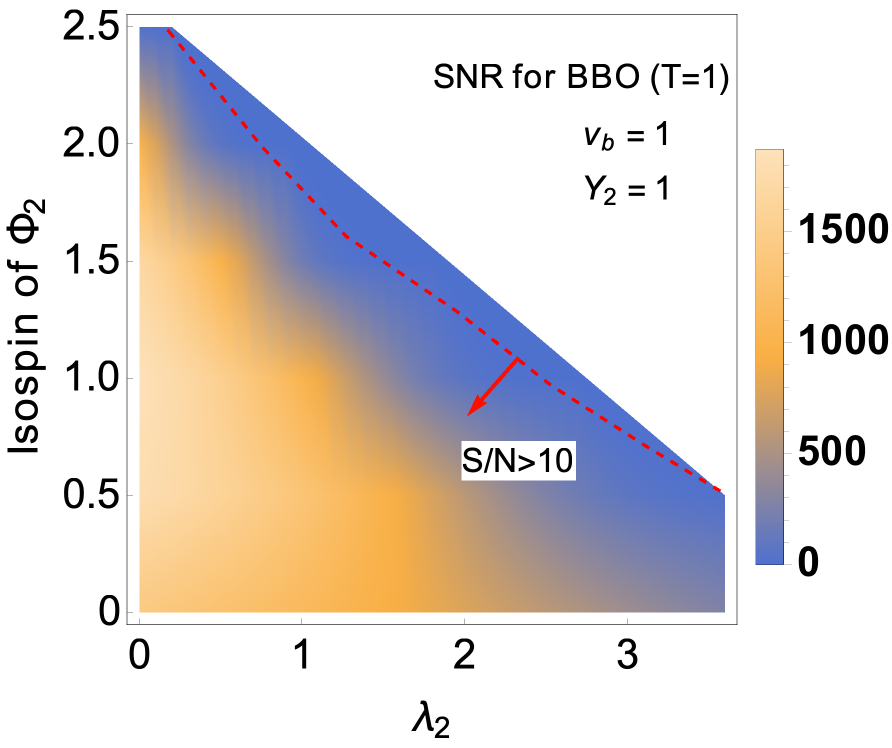}\\
\caption{The values of SNR for the model with CSI at LISA, DECIGO and BBO. The experimental period is one year and bubble wall velocity is 1. The red dashed line represents SNR=10.}
\label{fig:SNCSI}
\end{center}
\end{figure*}

In this section, we first show the value of $\phi_C/T_C$ for the model I in the first panel of Fig.~\ref{fig:PTGWCSI} to see which part of parameter space could produce detectable GWs.
The horizontal axis is the quartic coupling $\lambda_2$ and the vertical axis is the isospin $I_{ \Phi_2}$ of the additional scalar boson fields. It is easy to see that the value of $\phi_C/T_C$ is large for either small isospin $I_{ \Phi_2}$ or small $\lambda_2$ coupling. This result of $\phi_C/T_C$ is different from our intuition since the strongly FOPT can be typically realized by adding a large number of additional scalar boson fields into the model, which corresponds to a large isospin $I_{ \Phi_2}$. In the model with CSI, the number of additional scalar boson fields is related to the Higgs boson mass via Eq.~\eqref{eq:mhCSI}. Then, the $\phi_C/T_C$ without ring diagram contribution is roughly estimated by
\begin{equation}
\frac{\phi_C}{T_C} = \frac{E}{\lambda_2} \propto  I_{ \Phi_2}^{1/4}.
\end{equation}
On the other hand, the ring diagram contribution has $I_{ \Phi_2}^2$-dependence in Eq.~\eqref{eq:FmasshH}, therefore, $\phi_C/T_C$ becomes small through large ring diagram contributions from large $I_{ \Phi_2}^2$ value.

The PT parameters $T_n$, $\alpha$ and $\beta/H$ are shown in the last three panels of Fig.~\ref{fig:PTGWCSI}, respectively.
From these parameters, we can describe the GW spectrum from the FOPT. The SNR for the testability of this model with CSI is shown in Fig.~\ref{fig:SNCSI} with respect to the future space-borne GW detectors LISA, DECIGO, and BBO.
When evaluating the SNR, we simply take the mission duration to be one year and the bubble wall velocity to be one. The red dashed curves in the panels of DECIGO and BBO represent SNR=10. In the parameter regions to the left of the red dashed curves, the SNR is larger than 10, therefore, these parameter regions could be tested at future GW detectors DECIGO and BBO. From these figures, most of parameter regions for the model with CSI could be tested by the DECIGO and BBO missions.

\subsection{The model without classical scale invariance}

\begin{figure}
\begin{center}
\includegraphics[width=0.4\textwidth]{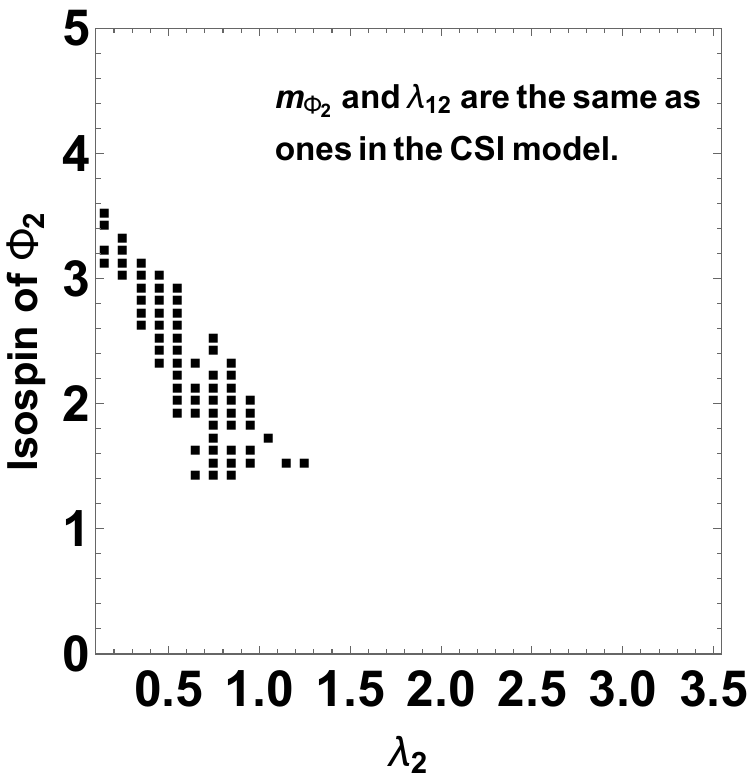}\\
\caption{The parameter region can have a stable minimum point along $\varphi_1$ and $\varphi_2$ axes.
At these black points, there is global minimum along $\varphi_2$ axis. }
\label{fig:multistep}
\end{center}
\end{figure}

\begin{figure*}
\begin{center}
\includegraphics[width=0.4\textwidth]{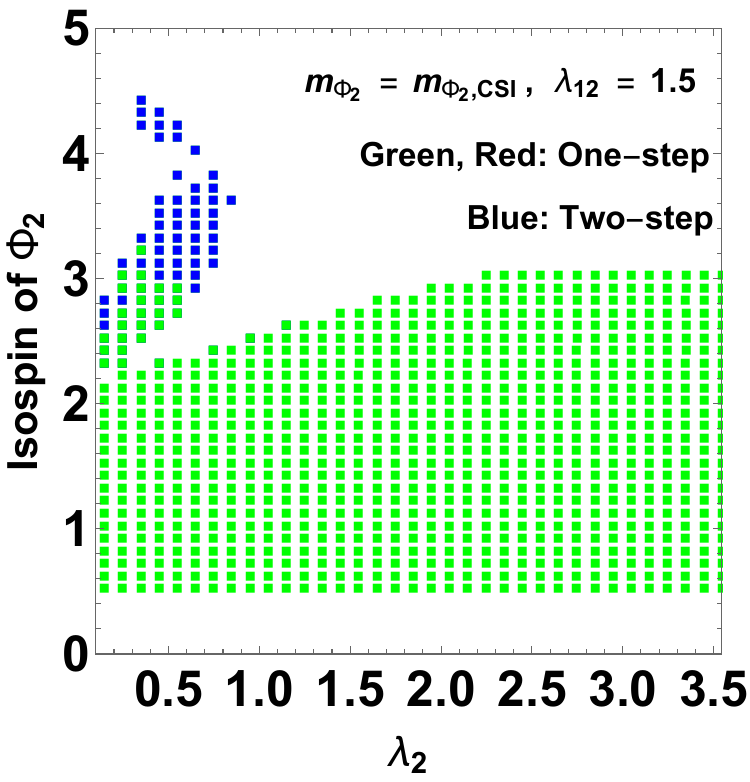}
\includegraphics[width=0.4\textwidth]{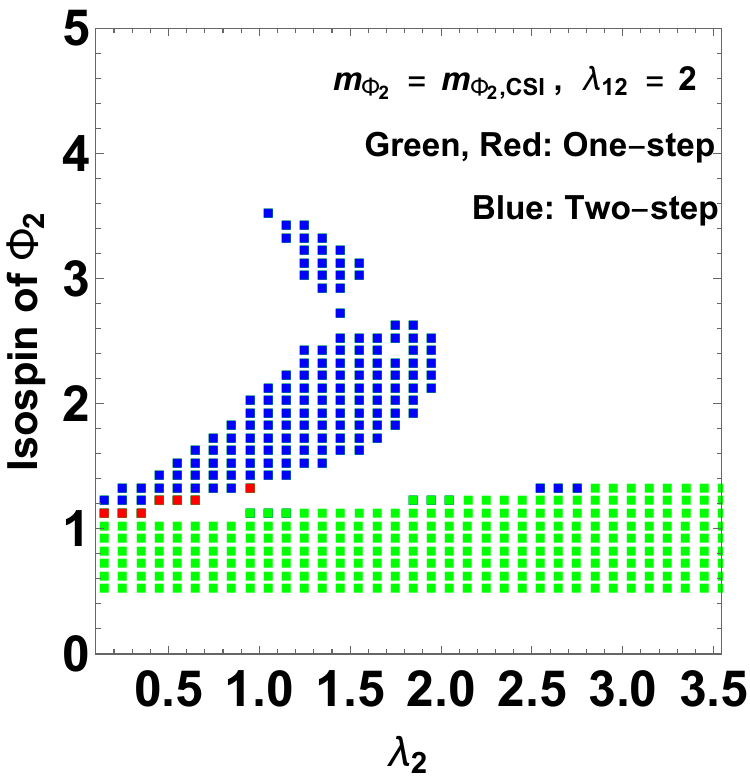}\\
\includegraphics[width=0.4\textwidth]{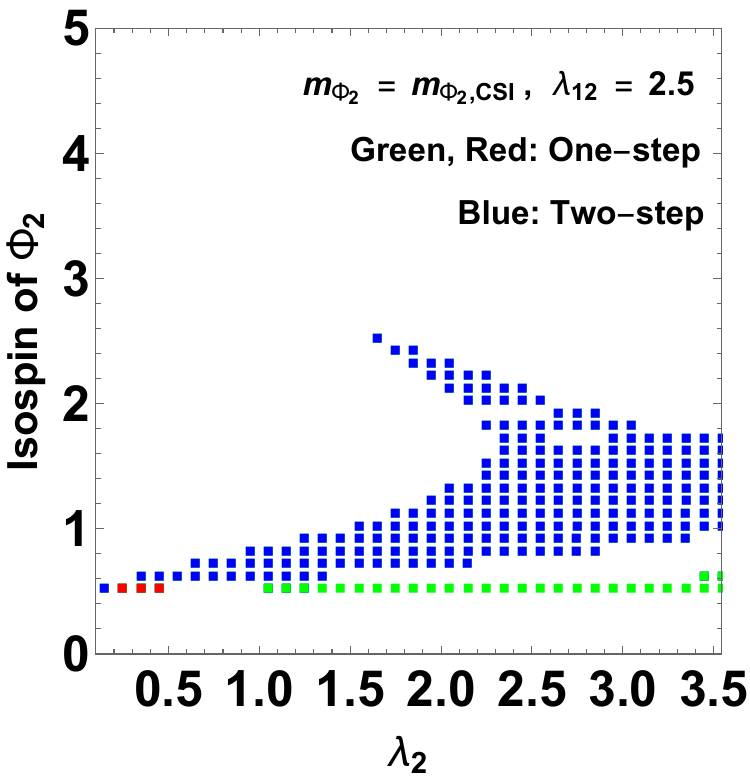}
\includegraphics[width=0.4\textwidth]{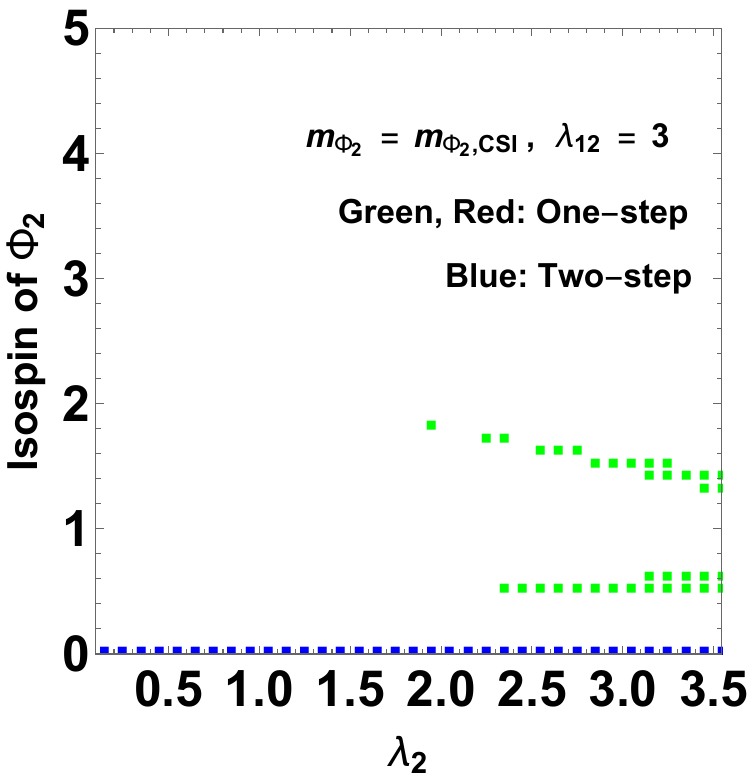}\\
\caption{Different PT regions in the parameter space supported by $\lambda_2$ and $\Phi_2$ for the model II with the same $m_{\Phi_2}$ as the model I but allowing for different $\lambda_{12}=1.5, 2.0, 2.5, 3.0$. The blue points represent the two-step PT.
The red and green points represent the one-step PT.
The detectable GWs at the BBO detection can be generated for the red points.
For the rest of regions in this figure, the global minimum point does not appear along  $\varphi_1$ axis otherwise the potential is unstable.}
\label{fig:1stPTwithoutCSI}
\end{center}
\end{figure*}

\begin{figure*}
\begin{center}
\includegraphics[width=0.4\textwidth]{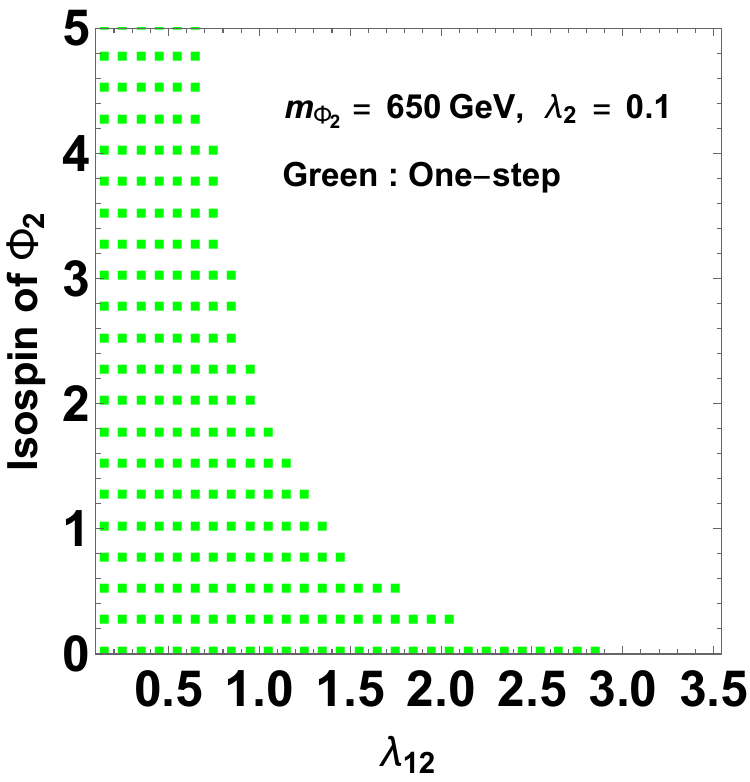}
\includegraphics[width=0.4\textwidth]{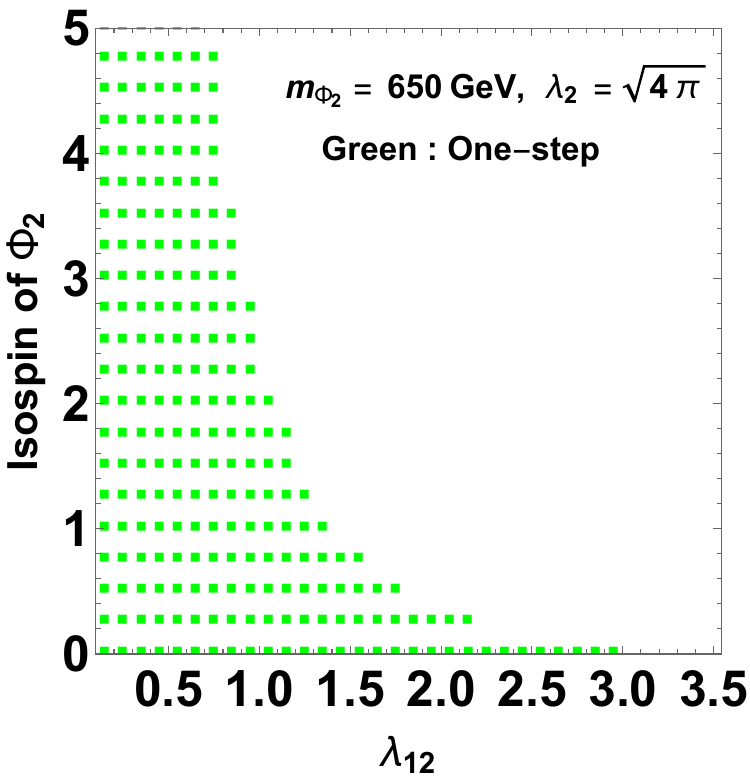}\\
\caption{The model parameters for the model II with $\lambda_{2}=0.1, \sqrt{4\pi}$ and $m_{\Phi_2}$ = 650 GeV.
The horizontal axis in left and right figures is $\lambda_{12}$. 
Otherwise, the same as Fig.~\ref{fig:1stPTwithoutCSI}.}
\label{fig:1stPTwithoutCSI2}
\end{center}
\end{figure*}

\begin{figure*}
\begin{center}
\includegraphics[width=0.3\textwidth]{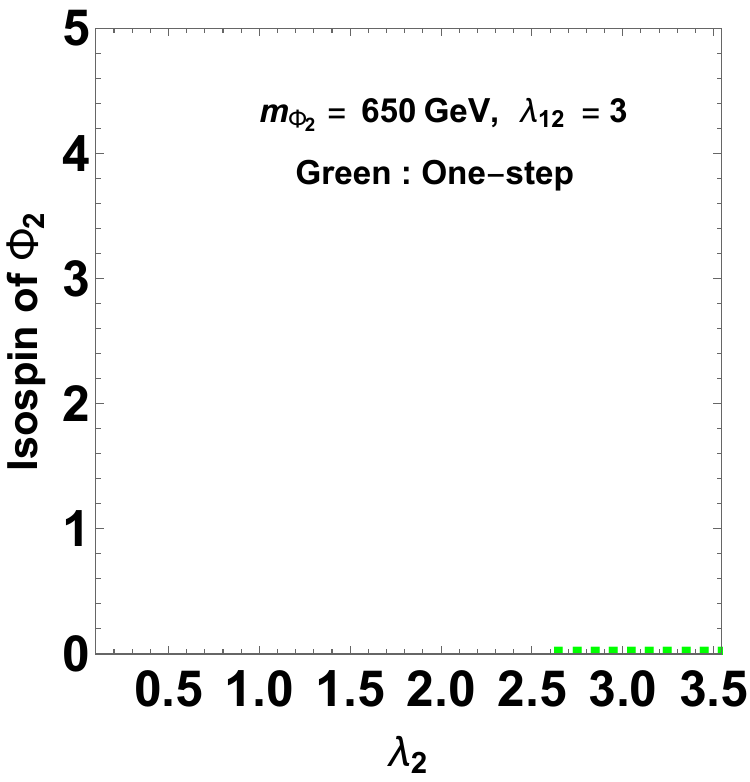}
\includegraphics[width=0.3\textwidth]{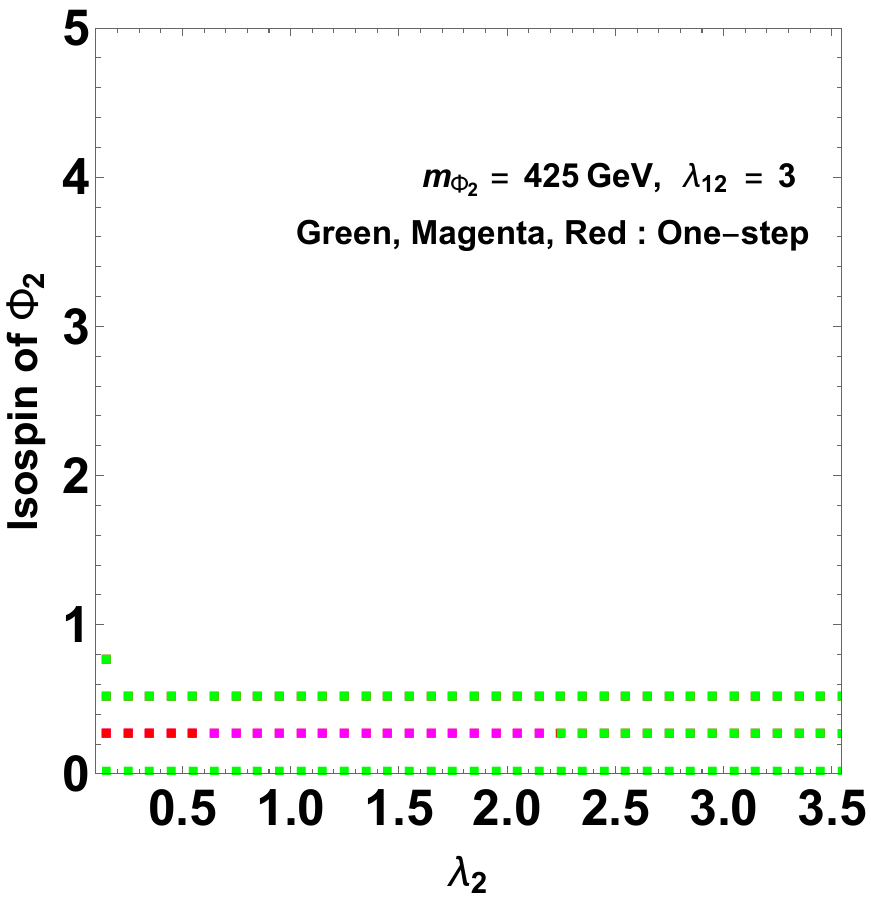}
\includegraphics[width=0.3\textwidth]{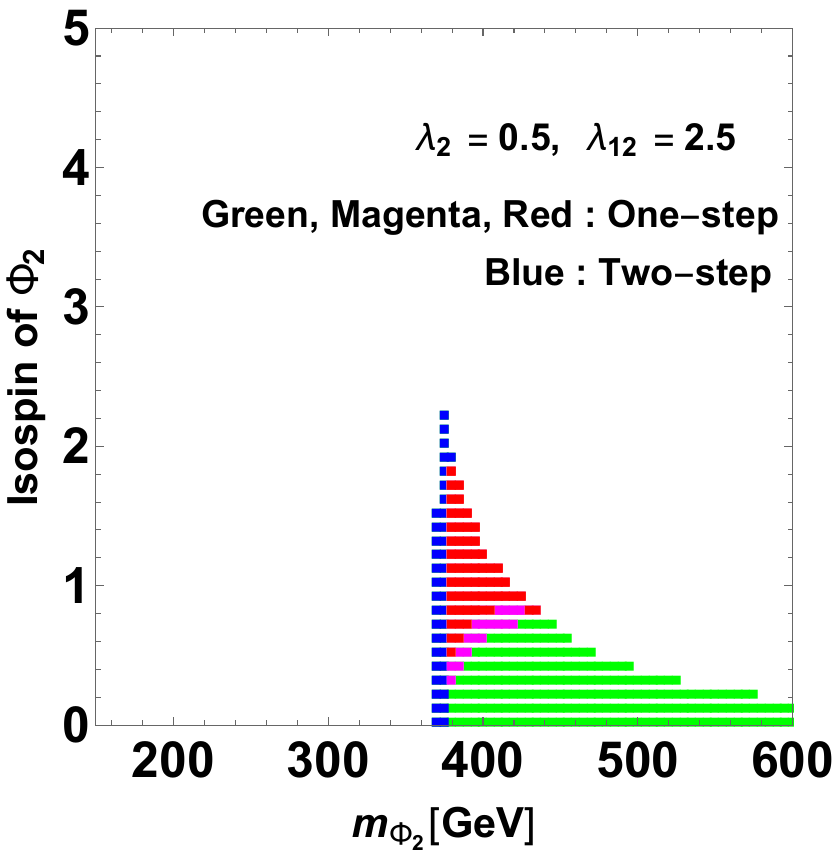}\\
\caption{(First two panels) The model parameters for the model II with $\lambda_{12}=$3 and $m_{\Phi_2}$ = 650, 425 GeV in the upper left and upper right panels, respectively.
At the red (magenta) marks, the detectable GWs at DECIGO (BBO) can be generated.
Otherwise, the same as Fig.~\ref{fig:1stPTwithoutCSI}.
(Last panel) The results in the model with $\lambda_{2}=$0.5 and $\lambda_{12}=$2.5.
Horizontal axis is the additional scalar boson mass.
The two-step PT occurs at blue marks.}
\label{fig:1stPTwithoutCSI3}
\end{center}
\end{figure*}

In this section, we show the results for the model II, which admits two more free parameters than the model I, namely, $m_{\Phi_2}$ and $\lambda_{12}$.
To compare with model I/III with only one-step PT (the additional VEV does not appear in the potential at zero temperature), we will focus on the parameter regions where the same one-step PT for model II as model I/III (green $\to$ red in Fig.~\ref{fig:Vtree}) can be realized.
We also neglect the parameter regions with other paths of PT, for example, two-step PT (namely the path along green $\to$ magenta $\to$ red, where the red is a global minimum).
We numerically examine the parameter regions where one-step EWPT can be generated by using following four parameter regions: (i) the same $m_{\Phi_2}$ and $\lambda_{12}$ as the model I, (ii) the same $m_{\Phi_2}$ as model I, (iii) the same $m_{\Phi_2}$ as model III, and (iv) different $m_{\Phi_2}$ and $\lambda_{12}$ from other models. 

In the parameter region (i), there are the minimum points along $\varphi_1$ and $\varphi_2$ axes at the black-point region in Fig.~\ref{fig:multistep}, but the minimum points along $\varphi_1$ and $\varphi_2$ axis are local and global minima, respectively.
According our numerical results, the two-step PT (green $\to$ red $\to$ magenta) is generated at the black-point region, and the magenta point becomes the true vacuum.
In such a case, we cannot explain the fermion mass since the additional scalar field $\Phi_2$ does not couple to the fermion fields anymore.
Therefore, in the parameter (i) for the model II, we cannot realize a proper PT which can make up the current universe. 
Although the model I can generate the detectable GW from the one-step EWPT in Fig.~\ref{fig:SNCSI}, the model II cannot realize the correct PT between green and red points in the parameter region (i), which shares the same parameter values as the model I. 
Therefore, we can distinguish the models with and without CSI by the GW detection.

In the parameter case (ii), model II admits the same $m_{\Phi_2}$ given by Eq.~\eqref{eq:MSCSI} as the model I but with $\lambda_{12}=1.5, 2.0, 2.5, 3.0$ decoupled from $I_{\Phi_2}$ by Eq.~\eqref{eq:lam12CSI} as shown in Fig.~\ref{fig:1stPTwithoutCSI}, respectively.
Note that in the case of model I,  $m_{\Phi_2}$ and $\lambda_{12}$ are related by $m_{\Phi_2}^2=\lambda_{12}v^2/2$, and thus we can get almost the same results for a similar case with the same $\lambda_{12}$ as model I by Eq.~(\ref{eq:lam12CSI}) but decoupled $m_{\Phi_2}$ from $I_{\Phi_2}$ by Eq.~\eqref{eq:MSCSI}.
All colored points in Fig.~\ref{fig:1stPTwithoutCSI} admit their global minimum along $\varphi_1$ (namely the red point in Fig.~\ref{fig:Vtree}). 
To show the parameter region where one-step PT along $\varphi_1$ axis can be generated, we compare the critical temperatures for the PT along $\varphi_1$ and $\varphi_2$ axes.
From the numerical results, the blue points in Fig.~\ref{fig:1stPTwithoutCSI} have higher critical temperature for the PTs along $\varphi_2$ than the PT for $\varphi_1$.
On the other hand, the green and red points in Fig.~\ref{fig:1stPTwithoutCSI} can realize the one-step PT (green $\to$ red in Fig.~\ref{fig:Vtree}), especially, the red points can realize the strong FOPT along $\varphi_1$ axis where the detectable GWs may be generated.
At these red points in the model with $\lambda_{12}=2$ and 2.5, the SNR for BBO with mission time $\mathcal{T}$=1 yr and $v_b=1$ is larger than 10.
Furthermore, the SNR for DECIGO with mission time $\mathcal{T}$=1 yr and $v_b=1$ at ($\lambda_{2}$, $I_{\Phi_2}$)=(0.25,0.5) is about 10.87, while other red points have their SNR no more than 10 for DECIGO.
In short summary, a small $\lambda_2$ is necessary for detectable GWs in the model II, which is not necessary for the model I as shown in Fig.~\ref{fig:SNCSI}.

To compare the results between the model II and the model III, we show the results in the parameter region (iii) with heavy $m_{\Phi_2}$ as model III.
For example, the results for heavy $m_{\Phi_2}=650$ GeV with some $\lambda_{2}$ values in the model II are shown in Fig.~\ref{fig:1stPTwithoutCSI2}, however, the effects from the additional heavy scalar field decouple.
From the additional scalar mass $M_{\Phi_2}(\varphi)^2\sim-\mu^2_2 + \lambda_{12}\varphi^2$ with large $\mu_2^2$ value, the cubic term of field-dependent mass in the effective potential can be expanded like $(M_{\Phi_2}^2)^{3/2}\sim|\mu^3_2|+\lambda_{12}\mu_2\varphi^2+ \lambda_{12}^2\varphi^4/\mu_2$, but the cubic $\varphi^3$ term does not appear in the potential.
On the other hand, the $M_{\Phi_2}(\varphi)^{3/2}$ with small $\mu_2^2$ value can have $\lambda_{12}^{3/2}\varphi^{3}$ as the source of a barrier.
Therefore, it is difficult to generate a sizable barrier to realize first-order EWPT in the model II with much heavy additional scalar field.
For the large negative $\mu_2^2$ term, the one-step PT occurs between green and red points in Fig.~\ref{fig:Vtree}.
In this case, $\lambda_2$, which is the coefficient of $\Phi_2^4$, does not much affect the EWPT, and the results are not very different between two figures of Fig.~\ref{fig:1stPTwithoutCSI2}.

The first two panels in Fig.~\ref{fig:1stPTwithoutCSI3} represent the results in the model with $\lambda_{12}$=3 and $m_{\Phi_2}=650$ and 425 GeV, which correspond to the parameter regions (iii) and (iv), respectively.
With these parameters, the mass parameter $\mu_2$ becomes small (non-decoupling) and the detectable GW spectrum can be generated.
Especially, in the model with $m_{\Phi_2}=$ 425 GeV, the red (magenta) marks can be described in the upper right figure, and the SNR for DECIGO (BBO) is larger than 10.
The last panel in Fig.~\ref{fig:1stPTwithoutCSI3} represents the results with respect to $I_{\Phi_2}$ and $m_{\Phi_2}$, which are related to parameter region (iv) in the model with benchmark parameters $\lambda_{2}=$0.5 and $\lambda_{12}=$2.5.
The blue marks in this model can realize two-step PT.
In short summary, the detectable GWs, which are represented by red and magenta marks, can be produced in the massive model with small isospin of $\Phi_2$ and small $\lambda_2$ value.
However, for the model II, such parameter regions with detectable GWs are not the same as those in model I and model III.
Therefore we may distinguish the three types of models by the GW observations.

\subsection{The model with dimension-six operator from $\Phi_2$}

\begin{figure*}
\begin{center}
\includegraphics[width=0.4\textwidth]{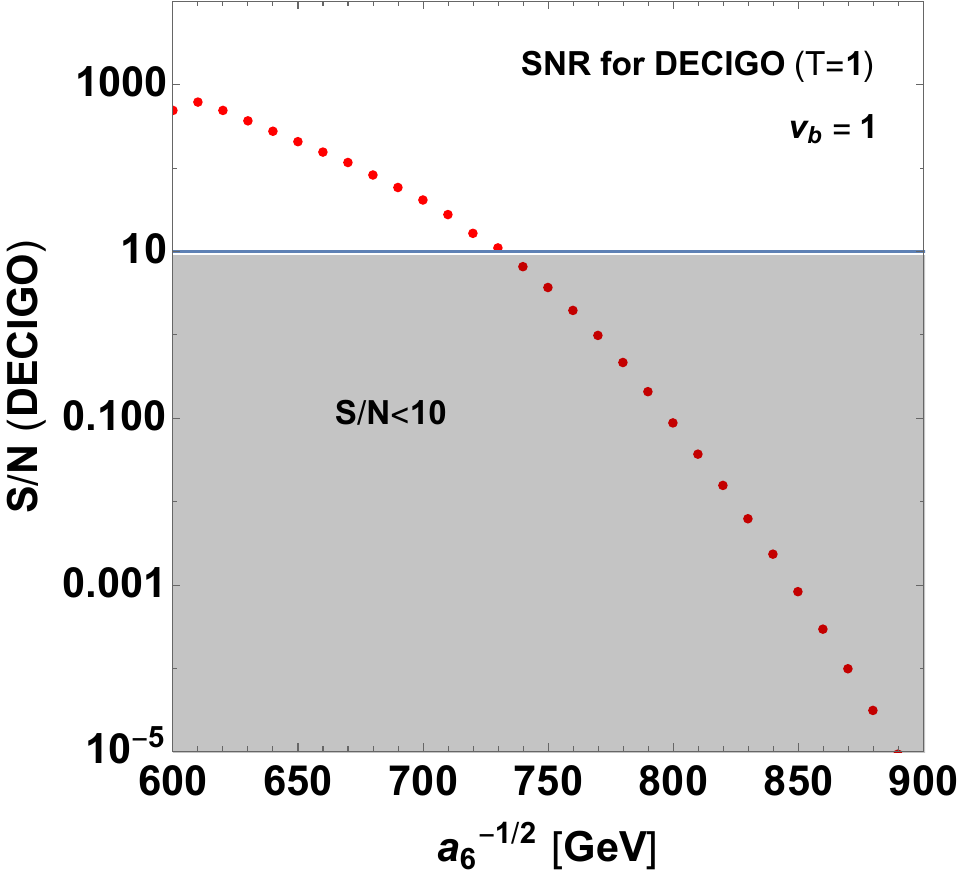}
\includegraphics[width=0.4\textwidth]{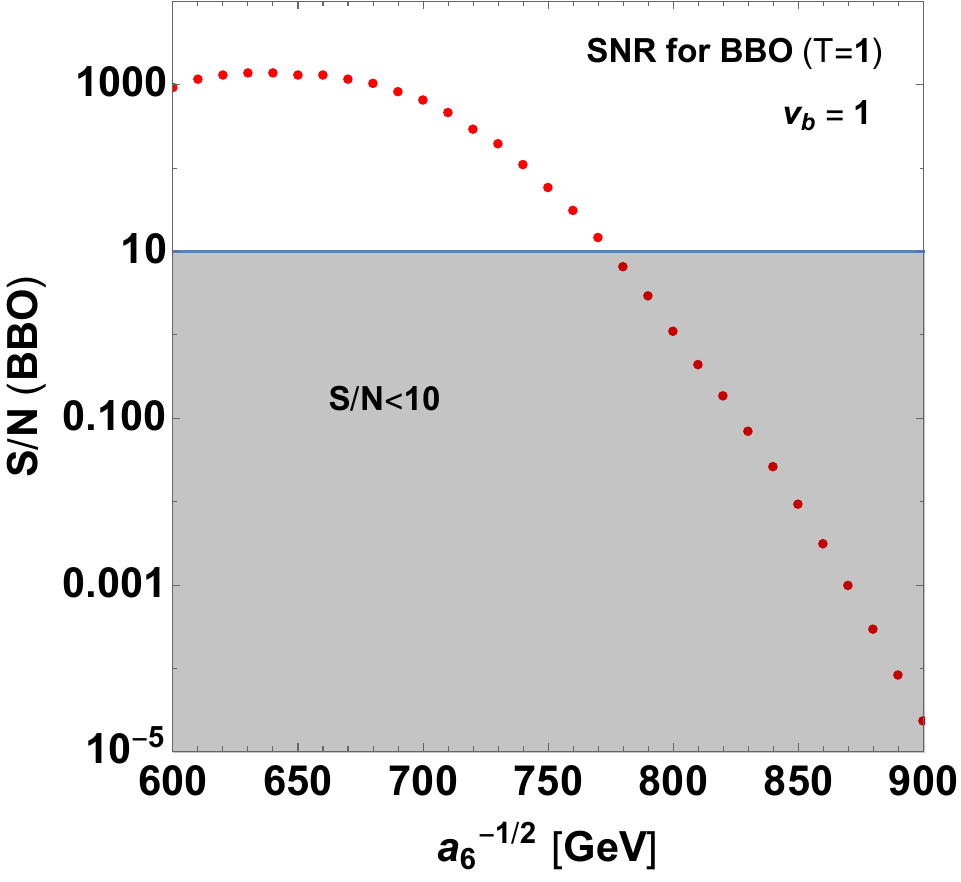}\\
\caption{The SNR value from the model with a higher dimensional operator for DECIGO and BBO with mission year $\mathcal{T}=1$ yr and $v_b=1$.}
\label{fig:simpledim6}
\end{center}
\end{figure*}

\begin{figure*}
\begin{center}
\includegraphics[width=0.4\textwidth]{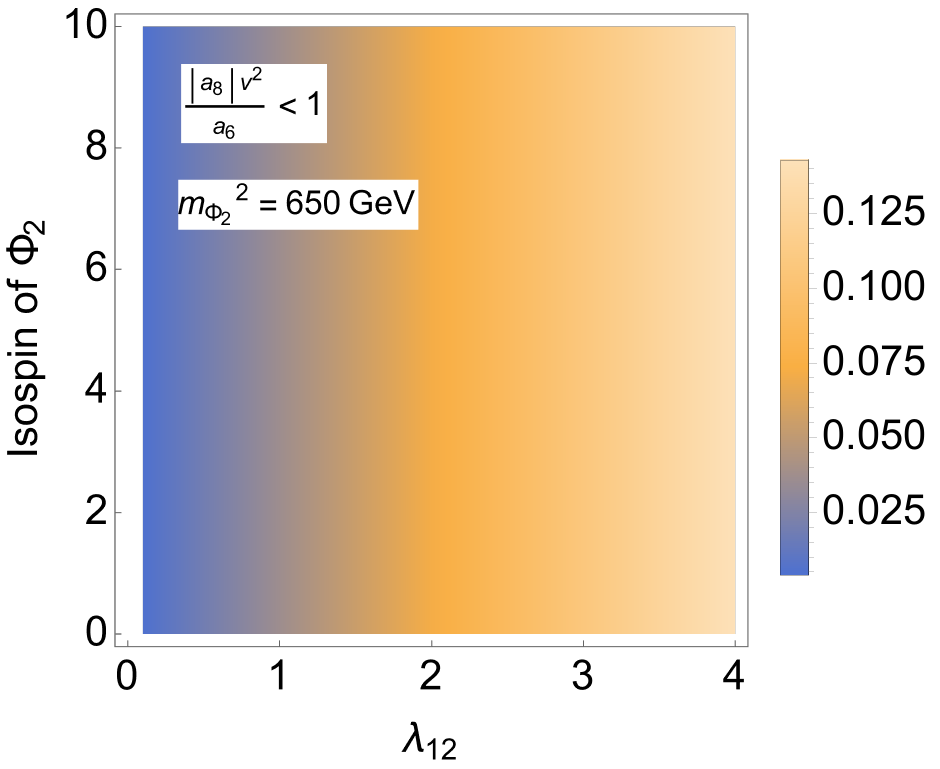}
\includegraphics[width=0.4\textwidth]{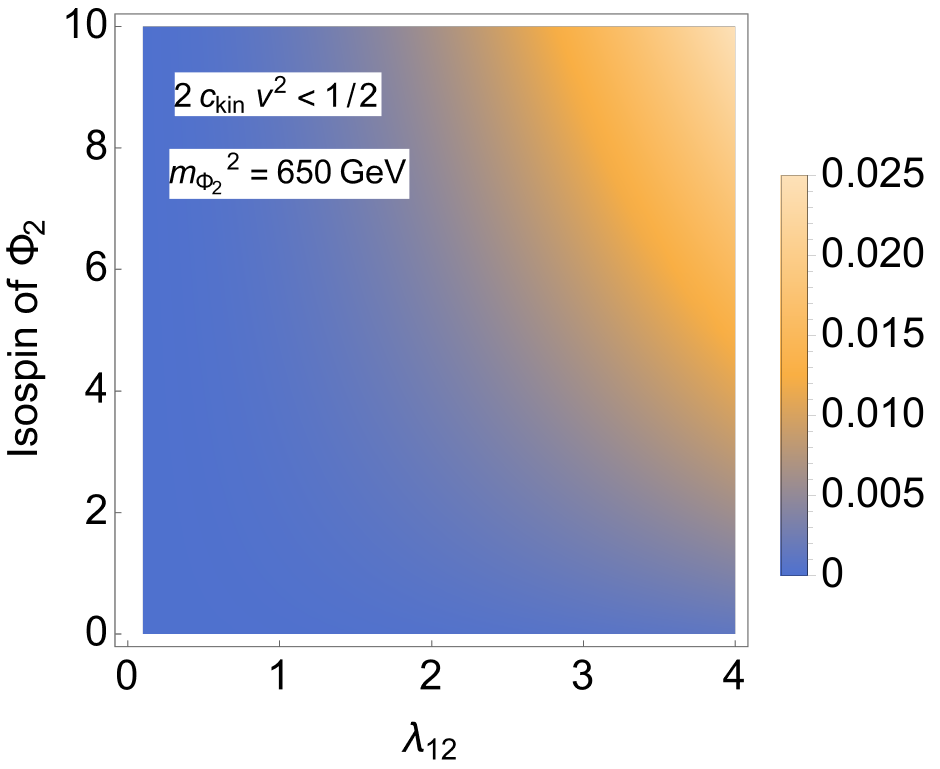}\\
\caption{The valid EFT expansion $2c_{kin}v^2<1/2$ and $|a_8|v^2/a_6<1$ in the model with a higher dimensional operator with respect to $I_{\Phi_2}$ and $\lambda_{12}$ and fixed $m_{\Phi_2}=650$ GeV.
Since the first condition comes from the comparison between one-loop dimension 6 and 8 operators, the values do not depend on the isospin $I_{\Phi_2}$. 
For the presented parameter regions, the conditions for a valid EFT expansion can be satisfied.}
\label{fig:EFTexp}
\end{center}
\end{figure*}

\begin{figure*}
  \begin{center}
\includegraphics[width=0.4\textwidth]{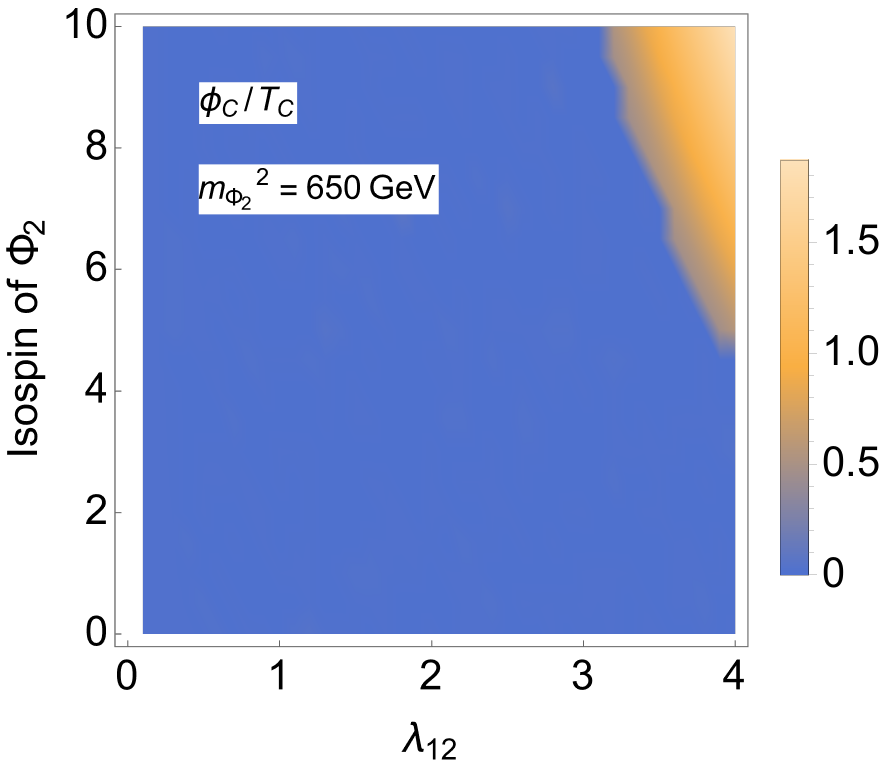}
\includegraphics[width=0.4\textwidth]{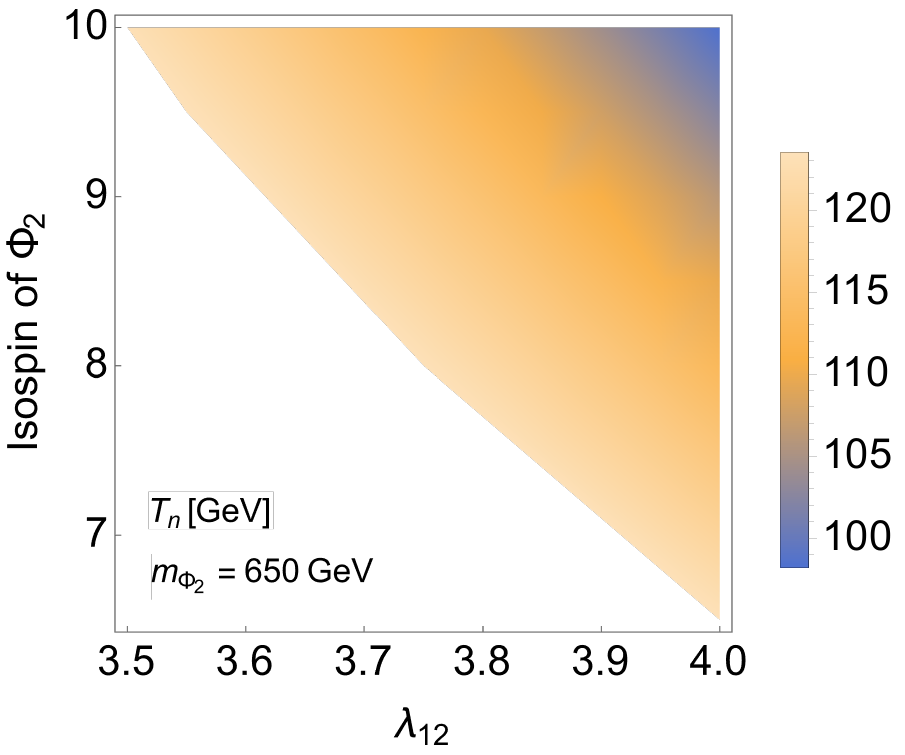}\\
\includegraphics[width=0.4\textwidth]{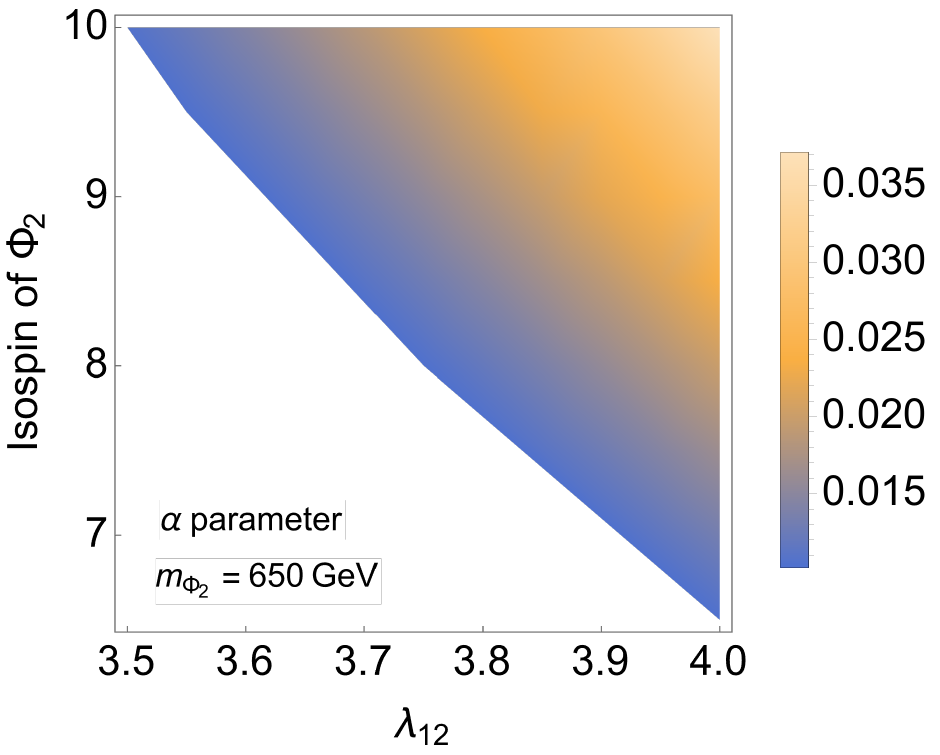}
\includegraphics[width=0.4\textwidth]{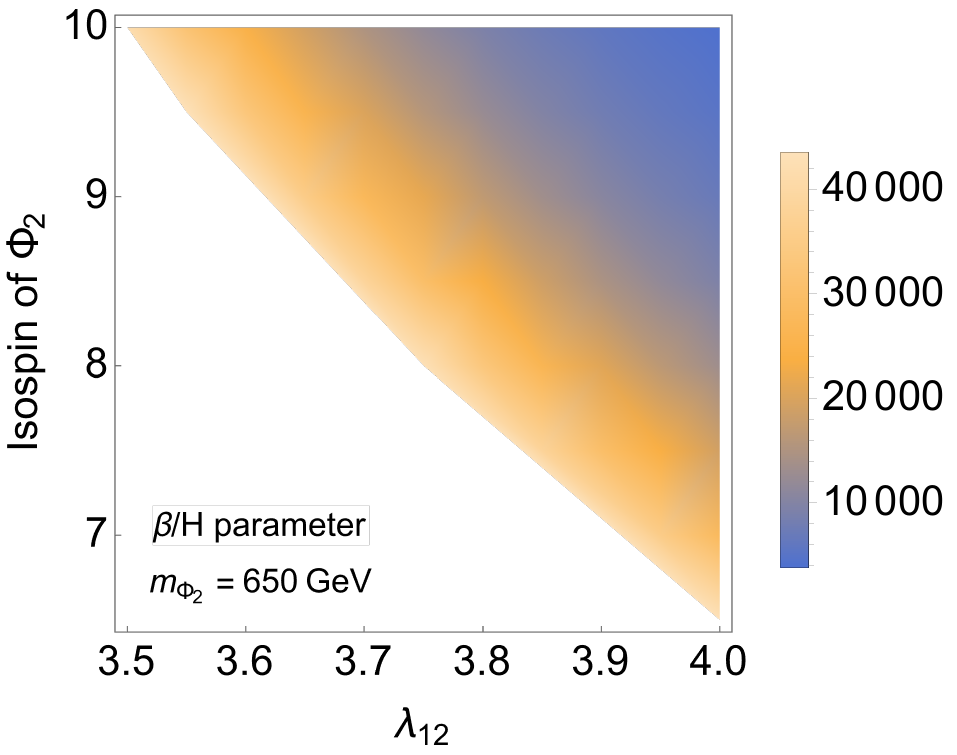}\\
\caption{The value of $\phi_C/T_C$, $T_n$, $\alpha$ and $\beta/H$ in the EFT model with $m_{\Phi_2} = 650$ GeV. Otherwise, the same as Fig.~\ref{fig:EFTexp}.}
\label{fig:1stPTEFT}
\end{center}
\end{figure*}

\begin{figure*}
\begin{center}
\includegraphics[width=0.4\textwidth]{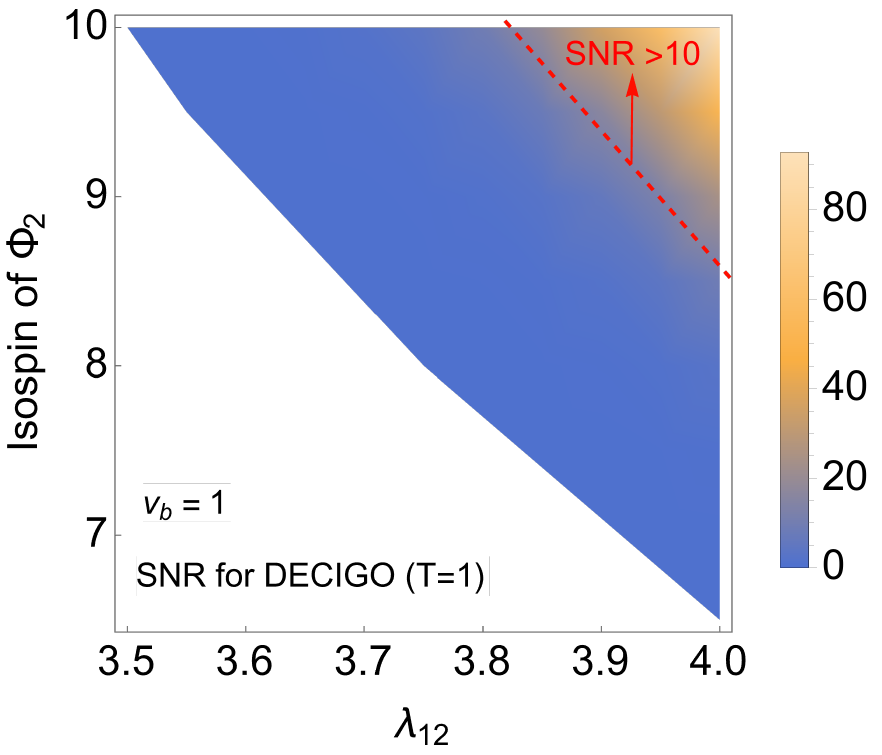}
\includegraphics[width=0.4\textwidth]{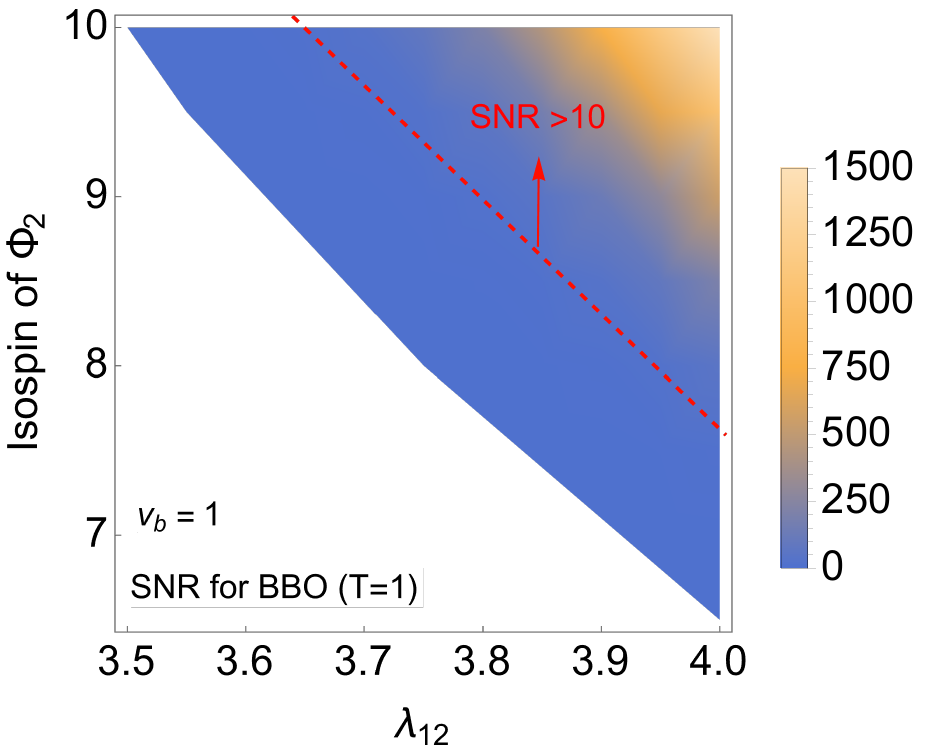}\\
\caption{The values of SNR in the EFT model with $m_{\Phi_2} = 650$ GeV at DECIGO and BBO. The ratio of SNR is larger than 10 in the parameter region to right of the red dashed lines. }
\label{fig:SNEFTmodel}
\end{center}
\end{figure*}

In this section, we discuss the testability for the model III.
At first, the parameter region of $a_6$ is shown in Fig.~\ref{fig:simpledim6} for detectable GWs at DECIGO and BBO with mission duration $\mathcal{T}=1$ yr and $v_b=1$.
In the dark regions, the SNR is smaller than 10, hence we require for a large $a_6$ value. 
Recall that with Eq.~(\ref{eq:EFTcoup}), $a_6=(1+2I_{\Phi_2})\frac{1}{(4\pi)^2m_{\Phi_2}^2}\left(\frac{\lambda_{12}^3}{8}+\frac{\lambda_{12}^2\lambda_1}{6} \right)$, a large $a_6$ value generally prefers a large isospin $I_{\Phi_2}$, namely a large number of the additional scalar boson fields. 
We should further check the conditions for a valid EFT expansion $|a_8|v^2/a_6<1$ and $2c_{kin}v^2<1/2$, which was discussed in Ref.~\cite{Postma:2020toi}.
In this model III, these conditions are 
\begin{equation}
\label{eq:EFTcond}
\frac{\lambda_{12}v^2}{2m_{\Phi_2}^2}<1,\quad (1+2I_{ \Phi_2})\frac{\lambda_{12}^2v^2}{12(4\pi)^2m_{\Phi_2}^2}<1 .
\end{equation}
The left and right panels of Fig.~\ref{fig:EFTexp} represents the values of the light hand side of these conditions.
The left panel is the comparison between dimension 6 and 8 operators.
Since the operators has one-loop level effects, the isospin $I_{ \Phi_2}$ cancel in this relation.
On the other hand, the right panel is the comparison between tree-level and one-loop level, therefore the value of this relation depends on the isospin $I_{ \Phi_2}$.
For the presented parameter space, the conditions for a valid EFT expansion can be satisfied.

The values of $\phi_C/T_C$, $T_n$, $\alpha$ and $\beta/H$ are then shown in Fig.~\ref{fig:1stPTEFT} and the corresponding SNR at DECIGO and BBO are shown in Fig.~\ref{fig:SNEFTmodel}, where the parameter regions to the right of red dashed lines have their SNR larger than 10. However, SNR is less than $10^{-5}$ for LISA, hence LISA may be difficult to test the parameter region of the model III. 
For detectable GWs from the model III, we require a large number of additional scalar fields (namely a large isospin $I_{\Phi_2}$) and large coupling $\lambda_{12}$, which can be distinguished from models I and II by detectable GWs in the parameter regions of small isospin $I_{\Phi_2}$. 
In particular, the model I cannot take the heavy scalar field with $m_{\Phi_2}=650$ GeV since the additional scalar mass should be less than 543 GeV from Eq.~(\ref{eq:MSCSI})~\cite{Hashino:2015nxa}.
Therefore, we may distinguish the model III from model I and model II by constraining the isospin of additional scalar fields with the GWs background observed at future GW detectors.

\section{Conclusions and discussions}\label{sec:condis}

The predictions for the GW background vary sensitively among different concrete models but also share a large degeneracy in the model buildings.
From that, in this time, we take into account an EFT treatment for three BSM models based on different patterns of the EWSB: (I) classical scale invariance, (II) generic scalar extension, and (III) higher dimensional operators.
In these EFTs, the EWSB can be realized by (I) radiative symmetry breaking, (II) Higgs mechanism, and  (III) EFT description of EWSB, respectively.

These three models can realize the strongly first-order EWPT and can produce the detectable GW spectrum by the effects summarized in Tab.~\ref{TAb}.
Here, $C_n$ are the effective couplings of the order parameter operators $\phi^n$ in the effective potential.
The dominant contributions of $\lambda_2$ and $I_{\Phi_2}$ in models (I) and (II) show up through the ring diagram effects as shown in Eqs~(\ref{withring1}), (\ref{piW}), (\ref{eq:FmasshH}) and (\ref{piWmassive}). 
Thus, a small $\lambda_2$ and a small $I_{\Phi_2}$ are required to produce the detectable GW spectrum.
The differences between models (I) and (II) are the $C_2$ and $C_4$ terms and the number of free parameters in the model.
In the model (II), these terms have the tree-level contribution, and thus we need to tune model parameters to have a sizable $C_3$ comparable with these $C_2$ and $C_4$ terms.
Unlike the model (I), we can use the $\lambda_{12}$ parameter to realize such a situation and could generate the first-order EWPT as shown in Fig.~\ref{fig:1stPTwithoutCSI}.
On the other hand, the $\lambda_2$ in model (III) does not contribute to the effective potential after integrating out the $\Phi_2$ scalar field, and the high dimensional operator $a_6^{-1/2}$ in Fig.~\ref{fig:simpledim6} is inversely proportional to $\lambda_{12}$ and $I_{\Phi_2}$ as shown in Eq.~(\ref{eq:EFTcoup}).
Therefore, a small $a_6^{-1/2}$ can be realized by a large $\lambda_{12}$ and a large $I_{\Phi_2}$.

\begin{table}[tp]
  \centering
  \begin{tabular}{|c|c|c|c|c|c|}
    \hline
    Model & $C_2\phi^2$&$C_3\phi^3$&$C_4\phi^4$&$C_6\phi^6$& GW features\\
       \hline\hline
     I   & Loop & Loop & Loop & None & Small $\lambda_2$ and small $I_{\Phi_2}$\\
     \hline
     II & Tree & Loop & Tree & None & Small $\lambda_2$ and small $I_{\Phi_2}$ \\
     \hline
      III & Tree & Loop & Tree & Tree & Large $\lambda_{12}$ and large $I_{\Phi_2}$ \\
    \hline
    \end{tabular}
      \caption{The potential forms in the three types of the models where the EWSB can be realized by (I) radiative symmetry breaking, (II) Higgs mechanism and (III) EFT description of EWSB, respectively.
      The last column shows the source of detectable GW spectrum in these models.
      Otherwise, the same as Tab.~\ref{TAB111}.}
  \label{TAb}
\end{table}

These three types of models might be distinguished by investigating the detectable GWs in the parameter regions with overlapping parameters:
(1) Model I and model II might be distinguished by the detection of GWs in the parameter regions as shown in Figs.~\ref{fig:SNCSI} and \ref{fig:multistep}. When taking the same $m_{\Phi_2}$ and $\lambda_{12}$ as model I by Eqs.~\eqref{eq:MSCSI} and \eqref{eq:lam12CSI}, model II cannot generate the correct one-step PT where the red point in Fig.~\ref{fig:Vtree} is the local minimum. However, when just taking the same $m_{\Phi_2}$ as model I by Eq.~\eqref{eq:MSCSI} but decoupling $\lambda_{12}$ from $I_{\Phi_2}$ by Eq.~\eqref{eq:lam12CSI}, model II could produce detactable GWs in the parameter regions of small $\lambda_2$ and small $I_{\Phi_2}$, which is partially overlapping with model I with detectable GWs. 
In this case, although we cannot fully distinguish models I and II by the GW detections alone, we may use other observations, such as $hhh$ coupling~\cite{Hashino:2016rvx} and $h\gamma\gamma$ coupling~\cite{Earl:2013jsa,Hashino:2015nxa}, to do that.
(2) Model II and model III can be distinguished by the detection of GWs in the parameter regions as shown in Figs.~\ref{fig:1stPTwithoutCSI2} and \ref{fig:SNEFTmodel} since detectable GWs are produced for model II in the parameter regions of small $\lambda_2$ and small $I_{\Phi_2}$, while model III favors a large $I_{\Phi_2}$ to generate detectable GWs. 
(3) Similarly, model I could also be distinguished from model III by detectable GWs in the parameter regions of small $I_{\Phi_2}$ (model I) and large $I_{\Phi_2}$ (model III). Furthermore, the additional scalar mass should be less than 543 GeV in the model I, different from the case of model III with a heavy scalar field.

Therefore, we may distinguish these three effective model descriptions of EWPT by future GW detections in space. Nevertheless, our PT analysis for the effective model descriptions is preliminary in surveying part of the parameter space, and it only depicts those scalar extensions (with $Z_2$ symmetry) of fundamental Higgs models \cite{Corbett:2017ieo,Li:2020gnx,Li:2020xlh} and Coleman-Weinberg Higgs models \cite{Hill:2014mqa,Helmboldt:2016mpi,Hashino:2015nxa}, but by no means covers all the effective models of EWPT. We will investigate the PT dynamics in a more general classifications of EWSB  \cite{Agrawal:2019bpm} in the ever-enlarging parameter space in future works.

\begin{acknowledgments}
This work is mainly supported by the National Key Research and Development Program of China Grant No. 2021YFC2203004, No. 2021YFA0718304, and No. 2020YFC2201501,
R. G. C. is supported by the National Natural Science Foundation of China Grants No. 11947302, No. 11991052, No. 11690022, No. 11821505 and No. 11851302, the Strategic Priority Research Program of the Chinese Academy of Sciences (CAS) Grant No.XDB23030100, No. XDA15020701, the Key Research Program of the CAS Grant No. XDPB15, the Key Research Program of Frontier Sciences of CAS.
S. J. W. is supported by the National Key Research and Development Program of China Grant No. 2021YFC2203004, No. 2021YFA0718304, the National Natural Science Foundation of China Grant No. 12422502, No. 12105344,  the China Manned Space Project with NO.CMS-CSST-2021-B01.
J. H. Y. is supported by the National Science Foundation of China under Grants No. 12022514, No. 11875003 and No. 12047503, and National Key Research and Development Program of China Grant No. 2020YFC2201501, No. 2021YFA0718304, and CAS Project for Young Scientists in Basic Research YSBR-006, the Key Research Program of the CAS Grant No. XDPB15.
\end{acknowledgments}

\appendix

\section{Stationary condition and CP-even boson masses in the model without CSI}\label{app:app1}

We use the stationary condition at ($\varphi_1$, $\varphi_2$) = ($v$, 0) and the second derivatives of the potential with one-loop effects,
\begin{align}
 \left. \frac{\partial V_{eff}}{\partial \left\langle \Phi_1 \right\rangle} \right|_{ \left\langle \Phi_1 \right\rangle =v, \left\langle \Phi_2 \right\rangle =0} = 0,\quad  \left. \frac{\partial^2 V_{eff}}{\partial \left\langle \Phi_1 \right\rangle^2} \right|_{ \left\langle \Phi_1 \right\rangle =v, \left\langle \Phi_2 \right\rangle =0} = m_h^2,
\end{align}
where we take $Q=v=246$ GeV. With tadpole condition ${\cal M}_{\Phi_2\Phi_2}^2>{\cal M}_{\Phi_1\Phi_1}^2$, one arrives at
\begin{widetext}
\begin{align}
\label{eq:redc1}
\left.\frac{\partial V_{\rm eff}}{\partial  \left\langle \Phi_1 \right\rangle} \right|_{ \left\langle \Phi_1 \right\rangle=v,  \left\langle \Phi_2 \right\rangle=0} &=
	-\mu_1^2+\lambda_1 v^2 +\frac{1}{16\pi^2} \Bigg[ \frac{6\lambda_1+\lambda_{12}}{4} f_+(m_{\Phi_{1,r}}^2,m_{\Phi_{2,r}}^2) + \frac{1}{4}   (6\lambda_1-\lambda_{12}) f_-(m_{\Phi_{1,r}}^2,m_{\Phi_{2,r}}^2) \nonumber\\
	& \quad +(2(2I_{ \Phi_2}-1)-1) \frac{\lambda_{12} m_{N_{2,r}}^2}{2}\left(\ln\frac{m_{N_{2,r}}^2}{Q^2}-1\right)  + \frac{6m_W^4}{ v^2}\left(\ln\frac{m_W^2}{Q^2}-\frac{1}{3}\right) \nonumber\\
	&\quad + \frac{3m_Z^4}{v^2}\left(\ln\frac{m_Z^2}{Q^2}-\frac{1}{3}\right) - \frac{12m_t^4}{v^2}\left(\ln\frac{m_t^2}{Q^2}-1\right)   \Bigg]= 0, 
\end{align}
\end{widetext}
where
\begin{align}
&f_\pm(m_i^2, m_j^2) = m_i^2\left(\ln\frac{m_i^2}{Q^2}-1\right)  \pm m_j^2\left(\ln\frac{m_j^2}{Q^2}-1\right), \\
&\Delta m_\Phi^2 = m_{\Phi_2}^2 - m_{\Phi_1}^2= \sqrt{\left({\cal M}_{\Phi_1\Phi_1}^2-{\cal M}_{\Phi_2\Phi_2}^2\right)^2+4{\cal M}_{\Phi_1\Phi_2}^4},\\
& \mathcal{M}_{neutral\,\,Higgs}^{2}=
	\left(
	\begin{array}{ccc}
{\cal M}_{\Phi_1\Phi_1}^2  & {\cal M}_{\Phi_1\Phi_2}^2 \\ 
{\cal M}_{\Phi_2\Phi_1}^2 & {\cal M}_{\Phi_2\Phi_2}^2
	\end{array}
	\right)
\end{align}
and the masses in the tadpole conditions of red point case are 
 \begin{align}
m_{\Phi_{1,r}}^2 = -\mu_1^2 + 3\lambda_1 v^2, &\quad m_{\Phi_{2,r}}^2 = -\mu_2^2 + \lambda_{12} v^2/2, \\
 m_{N_{1,r}}^2 = -\mu_1^2 + \lambda_{1} v^2, &\quad  m_{N_{2,r}}^2 = -\mu_2^2 + \lambda_{12} v^2/2, 
 \end{align}
with $m_{\Phi_{2}}^2>m_{h}^2$. In order to replace two of the input parameters in the potential in terms of the Higgs boson mass $m_h$ and additional neutral CP-even boson mass $m_{ \Phi_2}$ \footnote{For simplicity, we ignore the contribution from the two-point-function $\Pi_{hh}$ of the Higgs boson.}, we use
\begin{widetext}
\begin{align}
\left.\frac{\partial^2 V_{\rm eff}}{\partial  \left\langle \Phi_1 \right\rangle^2}  \right|_{ \left\langle \Phi_1 \right\rangle=v,  \left\langle \Phi_2 \right\rangle=0}  &=
	2 \lambda_1v^2 +\frac{v^2}{32\pi^2} \Bigg[A^r_2\ln\frac{m_{\Phi_2}^2m_h^2}{Q^4} -A^r_3\ln\frac{m_h^2}{m_{\Phi_2}^2}+  4\lambda_{12}^2(2(2I_{ \Phi_2}-1)-1) \ln\frac{m_{N_{2,r}}^2}{Q^2} \nonumber\\
&\quad + 12  \frac{m_Z^4}{v^4}\left(\ln\frac{m_Z^2}{Q^2}+\frac{2}{3}\right) +24  \frac{m_W^4}{ v^4}\left(\ln\frac{m_W^2}{Q^2}+\frac{2}{3}\right)  -48  \frac{m_t^4}{ v^4}\ln\frac{m_t^2}{Q^2} \Bigg]\equiv m_h^2 \label{eq:redc3}  \\
\left.\frac{\partial^2 V_{\rm eff}}{\partial  \left\langle \Phi_2 \right\rangle^2} \right|_{ \left\langle \Phi_1 \right\rangle=v,  \left\langle \Phi_2 \right\rangle=0} &= -\mu_2^2 +  \lambda_{12}v^2 +\frac{1}{64\pi^2} \Bigg[ B_1f_-(m_{\Phi_2}^2,m_h^2)+ B_2f_+(m_{\Phi_2}^2,m_h^2) \nonumber\\
&\quad  + 4(2(2I_{ \Phi_2}-1)-1) \lambda_2 m_{N_{2,r}}^2 \left(\ln\frac{m_{N_{2,r}}^2}{Q^2} - 1\right) + 12  \frac{m_Z^4Y_{ \Phi_2}^2}{v^2}\left(\ln\frac{m_Z^2}{Q^2}-\frac{1}{3}\right)  \nonumber\\
&\quad +24  \frac{m_W^4I_W^2}{ v^2}\left(\ln\frac{m_W^2}{Q^2}-\frac{1}{3}\right) \Bigg]\equiv m_{ \Phi_2}^2,  \label{eq:redc4}
\end{align}
\end{widetext}
where
\begin{align}
  A^r_2 &=\frac{1}{4}\left( (6\lambda_1+\lambda_{12})^2 +(6\lambda_1-\lambda_{12})^2\right), \\
   A^r_3 &= -\frac{36\lambda_1^2-\lambda_{12}^2}{2}\\
   B_1&= \frac{2}{\Delta m^2}
\left[\left(-6\lambda_2+\lambda_{12}\right)  \left({\cal M}_{\Phi_1\Phi_1}^2-{\cal M}_{\Phi_2\Phi_2}^2\right)+ 4\lambda_{12}^2 v^2 \right]\\
B_2 &= 6\lambda_2+\lambda_{12} .
 \end{align}
By using Eqs.~\eqref{eq:redc1}, \eqref{eq:redc3} and \eqref{eq:redc4}, we can replace the model parameters: 
\begin{align}
  (\mu_1^2, \mu_2^2,  \lambda_1,  \lambda_2,  \lambda_{12} )\to (v, m_{ \Phi_2}^2 , m_h ,  \lambda_2,  \lambda_{12}).
\end{align}

\section{Landau pole in the model with CSI}

\begin{figure}
\begin{center}
\includegraphics[width=0.4\textwidth]{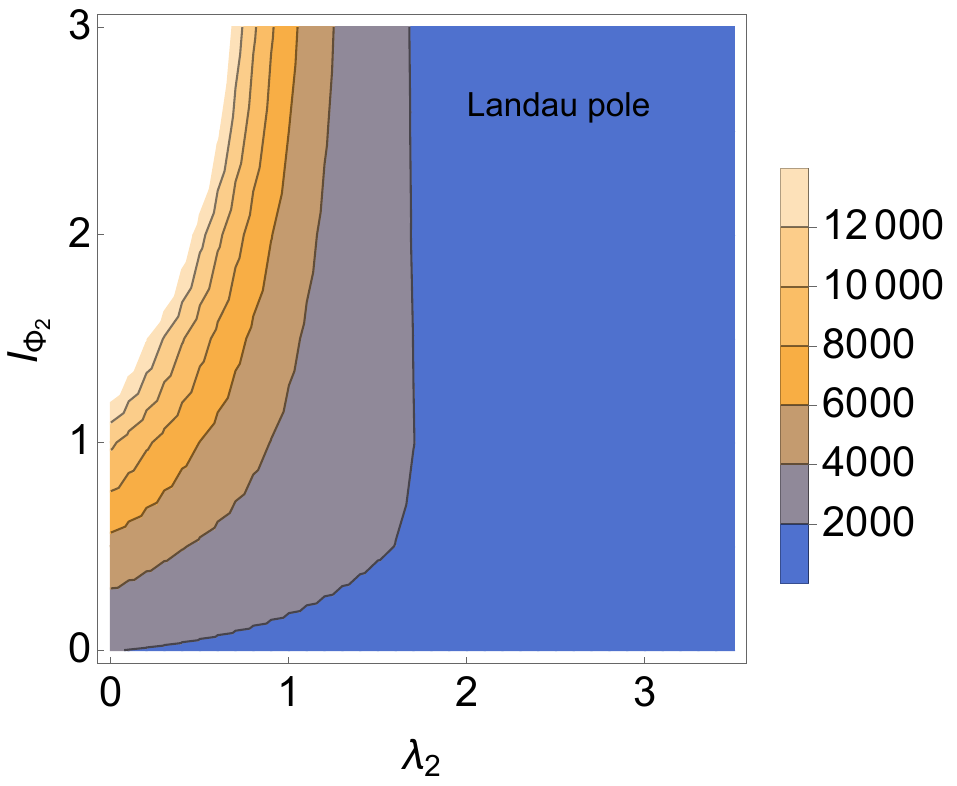}
\caption{The values of Landau pole in the CSI model.}
\label{fig:Landau}
\end{center}
\end{figure}

In this section, we show the Landau pole in the model with CSI.
The model with CSI typically has the large values of couplings, which can be obtained by Eq.~(\ref{eq:lam12CSI}).
The results of the Landau pole is given by Fig.~\ref{fig:Landau}
According to Figs.~\ref{fig:SNCSI} and \ref{fig:Landau}, the parameter region with detectable GW spectrum has Landau pole, which is O(1 TeV).
When the value of $\lambda_2$ will be large, the Landau pole also will be small.

\bibliography{ref}

\end{document}